\documentclass[twocolumn,showpacs,preprintnumbers,amsmath,amssymb,aps]{revtex4}
\usepackage{graphicx}% Include figure files
\usepackage{dcolumn}% Align table columns on decimal point
\usepackage{bm}% bold math
\usepackage{float}
\usepackage{soul}
\begin{document}

\title{Quantum coherence rather than quantum correlations reflect the effects of reservoir on the system's work capability}

\author{Hai Li$^{1,2}$}%
\author{Jian Zou$^{1}$}%
\email{zoujian@bit.edu.cn}
\author{Wen-Li Yu$^{3}$}%
 \author{Bao-Ming Xu$^{1}$}
 \author{Jun-Gang Li$^{1}$}%
\author{Bin Shao$^{1}$}%

 \affiliation{$^{1}$School of Physics, Beijing Institute of Technology, Beijing 100081, China}%
 \affiliation{$^{2}$School of Information and Electronic Engineering, Shandong Institute of Business and Technology, Yantai 264000, China}%
  \affiliation{$^{3}$School of Computer Science and Technology, Shandong Institute of Business and Technology, Yantai 264000, China}%
\date{\today}

%-----------------------abstract------------------------------------
\begin{abstract}

We consider a model of an optical cavity with a nonequilibrium
reservoir consisting of a beam of identical two-level atom pairs
(TLAPs) in the general X-state. We find that coherence of
multiparticle nonequilibrium reservoir plays a central role on the
potential work capability of cavity. We show that no matter whether
there are quantum correlations in each TLAP (including quantum
entanglement and quantum discord) or not the coherence of the TLAPs
has an effect on the work capability of the cavity. Additionally,
constructive and destructive interferences could be induced to
influence the work capability of cavity only by adjusting the
relative phase with which quantum correlations have nothing to do.
In this paper, the coherence of reservoir rather than the quantum
correlations effectively reflecting the effects of reservoir on the
system's work capability is demonstrated clearly.

%%increment
\end{abstract}

 \maketitle

\section{Introduction}\label{Sec:1}

Quantum coherence as a physical resource, being at the heart of
quantum interference has a variety of manifestations in different
areas of physics, and arises in some form or other in almost all the
phenomena of quantum mechanics and its applications~\cite{Ficek}.
The theoretical and experimental exploration of quantum coherence
has become a fascinating research topic. Recently, many
interesting investigations have been performed in various systems and
models, such as quantum optical systems including microwave
cavities~\cite{Scully,Quan2006,Sun,Kim,Pielawa,Sarlette}, ion
traps~\cite{Poyatos,Myatt}, optical lattices~\cite{Diehl,Wolf},
optomechanical systems~\cite{Clerk}, and biological
systems~\cite{Engel,Pani,Huelga,Wu1,Wu2,Kassal}. Meanwhile, it has
been suggested that coherent quantum dynamics can play an important
role in the initial steps of photobiological
processes~\cite{Ishizaki,Lee1,Cheng1,Collini1}.

Recently, the thermodynamic effects of quantum coherence have
attracted much attention and have been investigated based on quantum
thermodynamics cycles. In this aspect, except for exploiting
thermodynamics resource of quantum mechanical working materials
\cite{Scully,Quan2006,Sun,Hormoz,
Correa1,Quan1,Wang1,Linden1,Grimsmo1,Ueda1,Kim1} with the help of
quantum engine and refrigerator models, the importance of reservoir
manipulation has also been very recently acknowledged in the context
of quantum thermodynamics: It has been recently demonstrated that
superefficient operation of quantum heat engines may be achieved,
e.g., by reservoir squeezing ~\cite{Huang1,Correa2} and
coherence~\cite{Scully,Quan2006,Sun} or using more general types of
non-equilibrium reservoirs~\cite{Abah1,Mehta,Lutz}. Especially, the
exploration of reservoir's coherence in quantum thermodynamics has
provoked great interest and the optical cavity model with a
nonequilibrium coherent reservoir has been considered. Compared with
the situation of noncoherent reservoir some novel features could be
exhibited such as the improvement on work extraction and efficiency
in the thermodynamic cycle~\cite{Scully,Quan2006}, and heating and
cooling of cavity~\cite{Sun}.

However, previous investigations have mainly focused on the case of
single-particle reservoirs with coherence (e.g., the single
two-level~\cite{Sun} or three-level~\cite{Quan2006,Scully} coherent
reservoirs). For two-particle or multi-particle reservoirs the
quantum effects of coherence on the work capability of system have
no related reports. Meanwhile, we also notice that based on a
photo-Carnot engine model similar to the one presented in
Ref.~\cite{Scully} the thermodynamic effect of quantum correlations
has been investigated in Ref.~\cite{Lutz}. They considered a beam of
thermally entangled pairs of two-level atoms as a heat reservoir,
and expressed the thermodynamic efficiency of the engine in terms of
quantum discord (QD) of the atomic pair. They also showed that
useful work could be extracted from quantum correlations, and
believed that quantum correlations of the atomic pair are a valuable
resource in quantum thermodynamics. However, for  multiparticle
systems the quantum correlations including quantum entanglement (QE)
and QD and quantum coherence may appear in systems simultaneously,
and they are closely related~\cite{Plenio}. Which one, quantum
coherence or quantum correlations from reservoir, is the good
physical quantity to effectively reflect the effects of reservoir on
the system's work capability on earth? This is what we mainly
concern about in this paper. It is noted that most recently some
interesting works have devoted to the explorations of thermodynamic
effects of quantum
correlations~\cite{Quan1,Wang1,Linden1,Grimsmo1,Ueda1,Kim1} and
quantum coherence~\cite{Scully1,Scully2,Nalbach,Dorfman} where the
quantum systems with quantum correlations or coherence are generally
served as the working substance of quantum engines or thermodynamic
cycles, i.e., the quantum correlations or the coherence come from
the working substance of quantum engines. However, in the present
paper, we are interested in the quantum effects of reservoir's
coherence and quantum correlations in quantum thermodynamics which
is very different from their works.

In this paper, we choose the usual micromaser model (e.g.,
Refs.~\cite{Scully3,Meystre,Scully4,Orszag,Meschede,Casagrande,Cresser,Cresser96,Bergou,Guerra,Brune,Rempe}),
to illustrate our idea. In our model as depicted in Sec.~\ref{Sec:2}
we consider a series of two-level atomic pairs (TLAPs) initially
prepared in the general X-state passing through a cavity. Here, it
is emphasized that for choosing our model, two major reasons are
taken into account. Firstly, in contrast to the previous
works~\cite{Scully,Quan2006,Sun} we take a series of TLAPs with
coherence as a reservoir instead of a single two-level~\cite{Sun} or
three-level~\cite{Scully,Quan2006} atom reservoir, and aim to
discuss the quantum effects of multi-particle reservoir's coherence
and quantum correlations on the work capability of cavity field.
Secondly, we want to know which one, coherence or quantum
correlation, plays the decisive role on the thermodynamic properties
of the cavity field although the quantum correlations are closely
related to the coherence in multi-particle systems. Meanwhile, it is
more meaningful for choosing the general X-state of the injected
TLAPs because they include a wide class of quantum states such as
the general W and GHZ states. In this paper, we find that no matter
whether there are quantum correlations or not the constructive and
destructive interferences could be induced to influence the
thermodynamic properties (such as the entropy and the average photon
number) of cavity only via adjusting the relative phase of the
TLAPs. In this paper, we show that it is the reservoir's coherence
rather than the quantum correlations that can be used to reflect the
effects of reservoir on the system's work capability effectively.
Furthermore, we also notice that it is proper to measure the
potential work capability of the cavity by using the entropy of
cavity rather than the average photon number, except that the cavity
is in thermal equilibrium, and in this case, although the average
photon number and the entropy are different physical quantities,
they have similar behavior and the average photon number could also
be used to describe the potential work capability of the cavity.

The paper is organized as follows. In Sec.~\ref{Sec:2}, we present
our model of a single-mode cavity field interacting with a series of
TLAPs injected randomly. A quantum master equation of the
single-mode cavity field is derived. In Sec.~\ref{Sec:3}, by
considering the TLAPs being in a general X-state we investigate the
dynamics of cavity field. We analyze the role of reservoir's
coherence and quantum correlations in detail numerically and
analytically, and show that the good physical quantity reflecting
the effects of reservoir on the system's work capability is the
quantum coherence not the quantum correlations in our model.
Finally, we summarize our paper with some discussions in
Sec.~\ref{Sec:4}.

%%~~~~~~~~~~~~~~~~~~~~~~~~~~~~~~~~~~~~~~~~~~~~~~~~~~~~~~~~~~~~~~~~~~~
%%~~~~~~~~~~~~~~~~~~~~~~~~~~~~~~~~~~~~~~~~~~~~~~~~~~~~~~~~~~~~~~~~~~~
\section{Cavity QED model and master equation}\label{Sec:2}

In this paper, we consider a QED model that contains a single-mode
cavity field and a nonequilibrium reservoir consisting of amount of
TLAPs. We respectively denote the two atoms in each TLAP as $A$ and
$B$, and assume that there is no interaction between them. When the
TLAPs are sent through the cavity at random as depicted in Fig.1,
each atom $A$ ($B$) interacts with the single mode cavity via a
resonant Jaynes-Cummings (JC) coupling. The Hamiltonian of system
can be described as
 \begin{equation}\label{Eq:inter}
 \begin{split}
\hat{H}=&\hat{H}_{at}+\hat{H}_{ca}+\hat{H}_{int},
 \end{split}
\end{equation}
where $\hat{H}_{at}=\hbar \omega \sum_{j=1}^{2}\hat{\sigma}_j^{z}$,
$\hat{H}_{ca}=\hbar \omega \hat{a}^{\dag}\hat{a}$ and
$\hat{H}_{int}=g\hbar\sum_{j=1}^{2}(\hat{a}\hat{\sigma}_j^{+}+\hat{\sigma}_j^{-}\hat{a}^{\dag})$
with $\hat{\sigma}_j^{\pm}=\frac{1}{2}( \hat{\sigma}_j^{x}\pm i
\hat{\sigma}_j^{y})$ are independently the Hamiltonian of the TLAPs,
cavity field and interaction between the TLAPs and the cavity; $g$
and $\omega$ are independently the coupling constant and the
transition frequency between the energy levels corresponding to
excited state $|e\rangle$ and ground state $|g\rangle$ of each
two-level atom, $A$ and $B$; $\hat{a}$ $(\hat{a}^{\dag})$ is the
annihilation (creation) operator of the cavity and satisfies the
commutation relation $[\hat{a},\hat{a}^{\dag}]=1$;
$\hat{\sigma}_j^{x,y,z}$ ($j = 1, 2$) are the usual Pauli operators.

%%%%%%%%%%%%%%%%%%%%%%%%%%%%%%%%%%%%%%%%%%%%%%%%%%%%%%%%%%%%%%%%%%%%

\begin{figure} \includegraphics[width=8 cm]{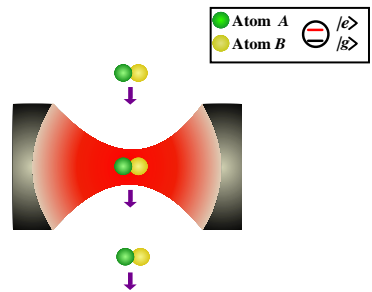}
\caption{(Color online). Schematic diagram of the dynamic model of a
series of TLAPs  initially prepared in a general X-state randomly
passing through a single-mode cavity. The two-level atoms in each
TLAP respectively denoted as $A$ (the left green solid ball) and $B$
(the right yellow one). When the TLAPs pass through the cavity the
coherence information in the TLAPs will transfer into the cavity.}
\label{fig1a}
\end{figure}
%%%%%%%%%%%%%%%%%%%%%%%%%%%%%%%%%%%%%%%%%%%%%%%%%%%%%%%%%%%%%%%%%%%%

We suppose that the pairwise TLAPs are randomly sent through the
cavity for a fixed time interval $\tau$ and there is at most one
TLAP in the cavity each time, and then the dynamic evolution of the
whole system (cavity + TLAP) during each time interval is a unitary
evolution and governed by the interaction Hamiltonian,
$\hat{H}_{int}$. The unitary evolution operator in the interaction
picture reads as
\begin{align}\label{unitary5}
\hat{U}(\tau )\equiv \exp(-i\hat{H}_{int}\tau)  = \left(
{\begin{array}{*{20}c}
 {\hat{ U}_{11}  } & {\hat{ U}_{12}  }  &  {\hat{ U}_{13}  }  & { \hat { U}_{14} }  \\
   {\hat{ U}_{21}  }  & {\hat{ U}_{22}  } &  {\hat{ U}_{23} } & {\hat{ U}_{24}  }   \\
   {\hat{ U}_{31} } & {\hat{ U}_{32} } &  {\hat{ U}_{33}  } & {\hat{ U}_{34} }  \\
   {\hat{ U}_{41} } & {\hat{ U}_{42} } &  {\hat{ U}_{43} } & {\hat{ U}_{44}   }
 \end{array}} \right),
 \end{align}
where the matrix elements are expressed as
\begin{equation}
\begin{aligned}
\hat U_{11}  =& 1 + 2\hat a\frac{{\hat A - 1}} { \hat\Lambda }\hat
a^{\dag},  ~~~~~ \hat U_{44}  = 1 + 2\hat a^{\dag}\frac{{\hat A -
1}}
{\hat\Lambda }\hat a,\\
\hat U_{22}  =& \hat U_{33}  = \frac{1} {2} ( {\hat A + 1}), ~~~~ \hat U_{23}  = \hat U_{32}  = \frac{1} {2} ( {\hat A - 1}),\\
\hat U_{14}  =& 2\hat a\frac{{\hat A - 1}} { \hat\Lambda }\hat a,~~~~~~~~~~~~~\hat U_{41}  =  2\hat a^{\dag}\frac{{\hat A - 1}} { \hat\Lambda }\hat a^{\dag},\\
\hat U_{12}  =& \hat U_{13}  =  - i\hat a\frac{{\hat
B}}{{\sqrt{\hat\Lambda}  }}, ~~~~~\hat U_{21}  = \hat U_{31}  =
-i\frac{{\hat B}}{{\sqrt {\hat\Lambda}  }}\hat a^ {\dag}   \hfill,\\
\hat U_{42}  =&\hat U_{43}  =  - i\hat a^ {\dag}  \frac{{\hat
B}}{{\sqrt {\hat\Lambda}  }},~~~\hat U_{24}  = \hat U_{34}  =  -
i\frac{{\hat B}}{{\sqrt {\hat\Lambda } }}\hat a, \hfill \nonumber
\end{aligned}
\end{equation}
and
\begin{equation}
\begin{aligned}
 \hat\Lambda=2({2\hat{a}^{\dag }\hat{a}}+1),~~\xi =g\tau,~~ &\hat{A}=\cos(\xi \sqrt{\hat\Lambda}),&\hat{B}= \sin (\xi
 \sqrt{\hat{\Lambda}}).\nonumber
\end{aligned}
\end{equation}
We assume that the $j$th TLAP is injected into the cavity at time
$t_j$, then, after a time interval $\tau $, the density matrix of
the cavity field becomes
\begin{equation}\label{timeEvolution}
\begin{aligned}
\hat{\rho}_{ca}(t_{j}+\tau)=\texttt{Tr}_{\textrm{AB}}[\hat{U}(\tau)
 \hat{\rho} _{\textrm{AB}} \otimes \hat{\rho}_{ca}(t_{j}) \hat{U}^{\dagger }\left( \tau \right)]
\equiv \mathcal{D}(\tau)\hat{\rho}_{ca}(t_{j}),
\end{aligned}
\end{equation}
where $\hat{\rho} _{\textrm{AB}}$ is the density matrix of the $j$th
TLAP and $\mathcal{D}(\tau)$ is a superoperator.

Since the TLAPs pass through the cavity randomly we assume that each
one arrives at the cavity with a probability $r$ per unit time. The
probability of a TLAP arrival, in a time interval of $(t,t+\delta
t)$, is $r\delta t$, and the probability without the TLAP passing is
$1-r\delta t$. Hereafter, for simplicity, we denote
$\hat{\rho}_{ca}$ as $\hat{\rho}$. Then we can obtain the density
matrix of cavity field at time $t+\delta t$ ~\cite{Sun}
\begin{equation}\label{densityMa8}
\begin{aligned}
\hat{\rho}(t+\delta t)=(1-r\delta t) \hat{\rho}(t) + r\delta
t\mathcal{D}(\tau)\hat{\rho}(t).
\end{aligned}
\end{equation}
For $\delta t\rightarrow 0$, one obtains the master
equation~\cite{Scully3,Orszag,Meschede,Casagrande,Cresser,Cresser96,Bergou,Guerra}
\begin{equation}
\begin{aligned}
\dot{\hat{\rho}}(t)=r[\mathcal{D}(\tau)-1]\hat{\rho}(t),\label{masterequation9}
\end{aligned}
\end{equation}
which describes the dynamics of the single-mode cavity field.

%%~~~~~~~~~~~~~~~~~~~~~~~~~~~~~~~~~~~~~~~~~~~~~~~~~~~~~~~~~~~~~~~~~~~~~~~~~~~~~~~~~~
%%~~~~~~~~~~~~~~~~~~~~~~~~~~~~~~~~~~~~~~~~~~~~~~~~~~~~~~~~~~~~~~~~~~~~~~~~~~~~~~~~~~~
\section {dynamics of cavity field with a nonequilibrium reservoir}{\label{Sec:3}

Here, we consider a nonequilibrium reservoir consisting of a beam of
TLAPs in a general X-state, and in the basis
$\{|ee\rangle,|eg\rangle,|ge\rangle,|gg\rangle\}$, the X-state is
given by
\begin{align}\label{Xstate}
\hat{\rho} _{AB}   = \left( {\begin{array}{*{20}c}
   { {a}_{11}  } & 0 &  0 & { {a}_{14} }  \\
   0 & { {a}_{22}  } &  { {a}_{23} } & 0  \\
   0 & { {a}_{32} } &  { {a}_{33}  } & 0  \\
   { {a}_{41} } & 0 &  0 & { {a}_{44}  }
 \end{array} } \right),
 \end{align}
where $\hat{\rho} _{AB}$ is normalized $\sum_{i=1}^4{a}_{ii}=1$, and
the nondiagonal elements ${a}_{14}={a}_{41}^*$ and
${a}_{23}={a}_{32}^*$. Inserting Eq. (\ref{Xstate}) into Eq.
(\ref{timeEvolution}) and after some calculations, the superoperator
$\mathcal{D}(\tau)$ can be expressed as
\begin{equation}\label{Evolution7}
\begin{aligned}
\mathcal{D}(\tau)\hat{\rho}_{}(t)=\sum_{i,j=1}^{4}{a}_{ij}\sum_{m=1}^4
\hat{U}_{mi}(\tau)\hat{\rho}_{}(t) \hat{U}_{mj}^{\dagger}(\tau),
\end{aligned}
\end{equation}
where ${a}_{ij}$ $(i,j=1,2,3,4)$ are given in Eq. (\ref{Xstate}).
Thus, for the X-state one has
\begin{equation}\label{Evolution10}
\begin{aligned}
\hat{\rho}(t_{j}+\tau)=  \sum_{i,j=1}^{4}{a}_{ij}\sum_{m=1}^4
\hat{U}_{mi}(\tau)\hat{\rho}_{}(t) \hat{U}_{mj}^{\dagger}(\tau)
\end{aligned}
\end{equation}
describing the density matrix of the cavity field at time $t_{j}+\tau$,
and the master equation Eq. (\ref {masterequation9}) can be
rewritten as
\begin{equation}\label{masterequation1}
\begin{aligned}
\dot{\hat{\rho}}(t)=r[\sum_{i,j=1}^{4}{a}_{ij}\sum_{m=1}^4
\hat{U}_{mi}(\tau)\hat{\rho}_{}(t)
\hat{U}_{mj}^{\dagger}(\tau)-\hat{\rho}(t)].
\end{aligned}
\end{equation}
 Eq (\ref{masterequation1}) is the quantum master equation of the cavity field
for each TLAP in the X-state. In order to obtain the dynamics of the cavity
field we consider two cases of the TLAPs passing through the cavity:
${1)}$ passing through instantly corresponding to $\xi \rightarrow
0$, and ${2)}$ passing through at a low speed corresponding to a
finite $\xi$. In case 1, Eqs. (\ref{Evolution10}) and
(\ref{masterequation1}) can be further expressed as
\begin{equation}
\begin{aligned}
\hat \rho ( {t_j + \tau })
 \approx&\xi ^2  \{ {{ (
{2a_{11}  + a _{22}  + a_{33}  + a_{23} + a _{32} }  ) [{\hat a^\dag
\hat \rho  ( {t_j })\hat{ a} -
\hat \rho  ( {t_j }  )}]} } \\
+ & ( {2a _{44} + a _{22} + a _{33}  + a _{23} + a_{32} } )\hat
a\hat \rho ( {t_j }  )\hat a^\dag
\\
-& ( {a _{11}  + a _{44} +a_{23} + a _{32}
} )\hat a^\dag \hat a\hat \rho  ( {t_j }  ) \\
-& ( {1 + a_{22}  + a_{33} + a_{23} + a _{32} } )\hat \rho ( {t_j }
)\hat a^\dag \hat a \\
+&a_{14} (2\hat{a}^{\dag }\hat{\rho} \hat{a}^{\dag }- \hat{a}^{\dag
}\hat{a}^{\dag }\hat{\rho}-\hat{\rho}\hat{a}^{\dag
}\hat{a}^{\dag }) \\
+&a_{41} (2\hat{a}\hat{\rho} \hat{a}- \hat{a}
\hat{a}\hat{\rho}-\hat{\rho}\hat{a}\hat{a})\}
 + \hat \rho ( {t_j } ) \label{evolutionequation18}
\end{aligned}
\end{equation}
and
\begin{equation}
\begin{aligned}
\dot{\hat{\rho}} \approx& r\xi^2\{a_{11} (2\hat{a}^{\dag }\hat{\rho}
\hat{a}-\hat{\rho} \hat{a}\hat{a}^{\dag
}-\hat{a}\hat{a}^{\dag }\hat{\rho}) \\
+&{a_{44} (2\hat{a}\hat{\rho} \hat{a} ^{\dag }-\hat{\rho}
\hat{a}^{\dag }\hat{a}-\hat{a}^{\dag }\hat{a}\hat{\rho}) }\\
-&(a_{22} +a_{33})[(2 \hat{\rho}\hat{a} ^{\dag
}\hat{a}-\hat{a}\hat{\rho}
\hat{a}^{\dag }-\hat{a}^{\dag }\hat{\rho}\hat{a})+\hat{\rho}] \\
+&a_{14} (2\hat{a}^{\dag }\hat{\rho} \hat{a}^{\dag }- \hat{a}^{\dag
}\hat{a}^{\dag }\hat{\rho}-\hat{\rho}\hat{a}^{\dag
}\hat{a}^{\dag }) \\
+&a_{41} (2\hat{a}\hat{\rho} \hat{a}- \hat{a} \hat{a}\hat{\rho}-\hat{\rho}\hat{a}\hat{a}) \\
+&(a_{23}  +a_{32} )(\hat{a}\hat{\rho}
\hat{a}^{\dag}+\hat{a}^{\dag}\hat{\rho}
\hat{a}-\hat{a}^{\dag}\hat{a}\hat{\rho}-\hat{\rho}\hat{a}^{\dag}\hat{a}-\hat{\rho})
\},\label{masterequation11}
\end{aligned}
\end{equation}
where we have made the approximation $\hat A=\cos (\xi \sqrt{{\hat
\Lambda }})\approx 1- \xi ^{2}{\hat\Lambda }/2$ and $\hat B =\sin
(\xi \sqrt{{\hat\Lambda }})\approx \xi \sqrt{{\hat\Lambda }}$, and
kept $\xi$ up to the second order. For simplicity, we denote
$\hat{\rho (t)}$ in Eq. (\ref{masterequation11}) as $\hat{\rho}$. In
case 2, the above approximation is not valid any more and from Eq.
(\ref{Evolution10}) the density matrix of cavity field at time
$t_{j}+\tau$ can be rewritten as
\begin{equation}
\begin{aligned}
\hat \rho(t_{j}+\tau)=&\sum_{m,n=0}^\infty \{\rho_{m,n}f_1
 +\rho_{m+1,n+1}f_2
+ \rho_{m-1,n-1}f_3 \\
+&\rho_{m-2,n-2}f_4 + \rho_{m+2,n+2}f_5 +\rho_{m,n+2}f_6\\
 +&\rho_{m+2,n}f_7
+ \rho_{m+1,n-1}f_8 +\rho_{m-1,n+1}f_9 \\
 +& \rho_{m,n-2}f_{10} +\rho_{m-2,n}f_{11} \} |m\rangle\langle n|,\label{evolution22}
\end{aligned}
\end{equation}
 where
$\rho_{m,-2}=\rho_{m,-1}=\rho_{-2,n}=\rho_{-1,n}=0$,
$(m,n=0,1,2,...)$, and during the derivation we have assumed that
the state of cavity field $\hat \rho(t_j)=\sum_{m,n=0}^\infty
\rho_{m,n}(t_j)|m\rangle\langle n|$ and denoted $\rho_{m,n}(t_j)$ as
$\rho_{m,n}$. For simplicity, $f_i$ ($i=1,2,3,...,11$) are given in
the appendix A. Though the expression of Eq. (\ref{evolution22}) is
complex, it, in the limit $\xi \rightarrow 0$, is in consistency
with Eq. (\ref{evolutionequation18}).

%%~~~~~~~~~~~~~~~~~~~~~~~~~~~~~~~~~~~~~~~~~~~~~~~~~~~~~~~~~~~~~~~~~~~~~~~~~~~~~~~~~~~~~~~~~~~~~~~~~~
%%% all the expressions of the evolution equation are given before;
%%Next, the dynamic parametres are derived and first for the <n> with the passing times
%%~~~~~~~~~~~~~~~~~~~~~~~~~~~~~~~~~~~~~~~~~~~~~~~~~~~~~~~~~~~~~~~~~~~~~~~~~~~~~~~~~~~~~~~~~~~~~~~~~~
Next, we first consider the dynamics of the cavity field in case 1. For
simplicity, throughout this paper we choose the vacuum state as the
initial state of the cavity field. From Eq.
(\ref{evolutionequation18}) it can be seen that the density matrices $\hat
\rho(t_{j}+\tau)$ and $\hat \rho(t_{j})$ possess the same form of
structure. When considering the cavity initially prepared in vacuum
state (being in the diagonal distribution), the density matrix of
cavity field, $\hat \rho(t_{j}+\tau)$, will keep the diagonal
distribution. Moreover, from Eq. (\ref{evolutionequation18}) we can
obtain that the average photon number of the cavity field at time ${t_j
+ \tau }$ reads as
\begin{equation}
\begin{aligned}
\langle \hat n(t_j  + \tau )\rangle=&\texttt{Tr}[ { \hat \rho} (t_j
+\tau )\hat n]\\
 = &[2\xi ^2 (a _{11}  -
a_{44} )+1] \langle\hat n(t_j ) \rangle \\
+ &\xi ^2 (2a _{11}  + a_{22} + a _{33}  + a _{23} + a_{32} ).
\label{photonnum16}
\end{aligned}
\end{equation}
From Eq. (\ref{photonnum16}) it can be seen that the increment of
average photon number between two neighboring passings $\Delta
\langle\hat n(t_j+\tau)\rangle=\langle\hat n(t_j + \tau ) \rangle -
\langle \hat n (t_j )\rangle$, satisfies
\begin{equation}
\begin{aligned}
\Delta \langle \hat n(t_j+\tau)\rangle
 =[1-2\xi ^2 (a_{44}  - a_{11} ) ]\Delta \langle
 \hat n(t_j) \rangle .\label{photonnum17}
\end{aligned}
\end{equation}
The increment ratio, $k$, for the two neighboring passings at time
$t_{j}$ and $t_{j}+\tau$ is directly obtained as
\begin{equation}
\begin{aligned}
k=\frac{\Delta \langle \hat n(t_j+\tau)\rangle}{\Delta \langle
 \hat n(t_j) \rangle}
= 1-2\xi ^2 (a_{44}  - a _{11} ). \label{photonnum18}
\end{aligned}
\end{equation}
Since the initial state of the cavity field is supposed to be the
vacuum state, i.e., $ \langle \hat n(0)\rangle=0$, from Eq.
(\ref{photonnum16}) the average photon number after the first
passing is
\begin{equation}
\begin{aligned}
\langle
 \hat n(\tau) \rangle
= \xi ^2 (2a_{11}  + a_{22} + a_{33}  + a_{23} + a_{32} ).
\label{firstphotonnum19}
\end{aligned}
\end{equation}
From Eqs. (\ref{photonnum18}) and (\ref{firstphotonnum19}) the
average photon number, after the $j$th time passing (i.e., at time
$t_j$), can be expressed as
\begin{equation}
\begin{aligned}
\langle \hat n(t_j)\rangle =\sum_{i=1}^{j}
 k^{i-1}\langle\hat n(\tau) \rangle. \label{firstphotonnum20}
\end{aligned}
\end{equation}
According to Eqs. (\ref{photonnum18}) and (\ref{firstphotonnum20})
for $a_{11}\geq a_{44}$ the ratio $k\geq1$ holds which means that
the average photon number, $\langle \hat n(t_j)\rangle$, is
divergent. On the contrary, for $k<1$ $(a_{11}< a_{44})$,
$\langle\hat n(t_j)\rangle$ is convergent, and in the limit
$j\rightarrow \infty$ (i.e., $t_j\rightarrow \infty$)
\begin{equation}
\begin{aligned}
\langle \hat n(t_j)\rangle|_{j\rightarrow \infty} = \frac{\langle
\hat n(\tau)\rangle}{1-k}
 = \frac{{2a _{11} + a_{22}  + a _{33}  + a_{23}  +
{a} _{32} }} {{2\left( {a _{44}  - a _{11} } \right)}}.
 \label{firstphotonnum21}
\end{aligned}
\end{equation}
 Eq. (\ref{firstphotonnum21}) shows that for fixed X-state
 with $a_{11}< a_{44}$ the cavity field, in the limit $t_j\rightarrow\infty$, can reach a
steady state.
%%~~~~~~~~~~~~~~~~~~~~~~~~~~~~~~~~~~~~~~~~~~~~~~~~~~~~~~~~~~~~~~~~~~~~~~~~~~~~~~~~~~~~~~~~~~~~~~~~~~
%%using the master equation to verify the consistence of the result of average photon valuve
%%~~~~~~~~~~~~~~~~~~~~~~~~~~~~~~~~~~~~~~~~~~~~~~~~~~~~~~~~~~~~~~~~~~~~~~~~~~~~~~~~~~~~~~~~~~~~~~~~~~
Besides, this result can also be verified by the master equation Eq.
(\ref{masterequation11}) from which, after some calculations, we
have
\begin{equation}
\begin{aligned}
\langle\dot {\hat n}\rangle = r\xi ^2 \left[ {2\left( {a _{11}  - a
_{44} } \right)\left\langle {\hat n} \right\rangle  + 2a_{11} + a
_{22}  + a _{33}  + a _{23}  + a _{32} } \right]=0,
\label{evolutionequation12}
\end{aligned}
\end{equation}
and
\begin{equation}\label{averphoton0}
\begin{aligned}
\left\langle { {\hat n}} \right\rangle _{ss} = \frac{{2a _{11} + a
_{22}  + a _{33}  + a _{23}  + {a} _{32} }} {{2\left( {a _{44}  - a
_{11} } \right)}},
\end{aligned}
\end{equation}
where $\langle \hat n \rangle_{ss}=\langle
 \hat n(t_j\rightarrow\infty) \rangle$ represents the average photon
number of the cavity in the steady state.

%%~~~~~~~~~~~~~~~~~~~~~~~~~~~~~~~~~~~~~~~~~~~~~~~~~~~~~~~~~~~~~~~~~~~~~~~~~~~~~~~~~~~~~~~~~~~~~~~~~~
%% explain that the classical correlation  has no effect on the work capability of the cavity field。
%%~~~~~~~~~~~~~~~~~~~~~~~~~~~~~~~~~~~~~~~~~~~~~~~~~~~~~~~~~~~~~~~~~~~~~~~~~~~~~~~~~~~~~~~~~~~~~~~~~~

Here, we consider two special noncoherent states of TLAPs,
$\hat{\rho} _{AB}^{(1)}$ with classical correlation and $\hat{\rho}
_{AB}^{(2)}$ without any correlation (i.e., product state), as
follows
\begin{equation}\label{cleq}
\begin{aligned}
{\hat{\rho}_{AB}^{(1)} }=&a_{11}| ee\rangle\langle ee|+a_{22}|eg
\rangle\langle eg|  +a_{33}|ge \rangle\langle ge|+a_{44}  |gg
\rangle\langle gg|,
\end{aligned}
\end{equation}
and
\begin{equation}\label{uceq}
\begin{aligned}
{\hat{\rho}_{AB}^{(2)} }=\hat{\rho}_{A}\otimes\hat\rho_B,
\end{aligned}
\end{equation}
where $\hat{\rho}_A$ and $\hat{\rho}_B$ respectively read as
\addtocounter{equation}{1}
\begin{align}
 \hat{\rho}_A= \left(
{\begin{array}{*{20}c}
 { a}_{11}+{ a}_{22}  &  0  \\
  0 & { a}_{33}+{ a}_{44}
 \end{array}} \right)  \tag{\theequation a}\\
 \hat{\rho}_B= \left(
{\begin{array}{*{20}c}
 { a}_{11}+{ a}_{33}  &  0   \\
  0 & { a}_{22}+{ a}_{44}
 \end{array}} \right). \tag{\theequation b}
\end{align}
It is noted that the choice of above two states enables the three
density matrices $\hat{\rho}_{AB}$ and $\hat{\rho}_{AB}^{(m)}$
(${m=1,2}$) to possess the same reduced density matrices
$\hat{\rho}_A$ and $\hat{\rho}_B$. In terms of Eq.
(\ref{averphoton0}) when the TLAPs are initially prepared in the states
$\hat{\rho}_{AB}^{(1)}$ and $\hat{\rho}_{AB}^{(2)}$, respectively,
one has
\begin{equation}
\begin{aligned}
\langle \hat n\rangle_{ss} ^{(m)}
 =\frac{{2a _{11} + a_{22}  + a _{33} }} {{2\left( {a _{44}  - a _{11} } \right)}},
 \label{avephot1}
\end{aligned}
\end{equation}
where $\langle \hat n\rangle_{ss} ^{(m)}$ represents the average
photon number of the cavity in the steady state for
$\hat{\rho}_{AB}^{(m)}$ (${m=1,2}$).

%%%~~~~~~~~~~~~~~~~~~~~~~~~~~~~~~~~~~~~~~~~~~~~~~~~~~~~~~~~~~~~~~~~~~~~~~~~~~~~~~~~~~~~~~~~~~~~~~~~~~
%%% explain that the quantum correlations are not the good quantity on the work capability of the cavity field。
%%%~~~~~~~~~~~~~~~~~~~~~~~~~~~~~~~~~~~~~~~~~~~~~~~~~~~~~~~~~~~~~~~~~~~~~~~~~~~~~~~~~~~~~~~~~~~~~~~~~~

%~~~~~~~~~~~~~~~~~~~~~~~~~~~~~~~~~~~~~~~~~~~~~~~~~~~~~~~~~

 Especially, if we consider
$\hat{\rho}_{A}=\hat{\rho}_{B}$ (i.e., $a_{22}=a_{33}$) and
$a_{11}+a_{22}<a_{33}+a_{44}$ the temperature of reservoir
consisting of TLAPs can be defined well by the two-level atoms. For
simplicity, we denote ${p}_{e}=a _{11} + a_{22}$ and ${p}_{g}=a
_{22} + a_{44}$ the inverse temperature of reservoir
$\beta_{\textrm{eff}}$ ($\beta_{\textrm{eff}}=1/(k_{B}T)$, $k_{B}$
is the Boltzmann constant) is expressed as
\begin{eqnarray} \label{bathtemp}
\frac{p_{e}}{p_{g}}=e^{-\beta_{\textrm{eff}}\omega}\Rightarrow\beta_{\textrm{eff}}=-\frac{1}{\omega}\ln\frac{p_{e}}{p_{g}},
\end{eqnarray}
where we let $\hbar =1$.
In this case, the asymptotic solution of the master equation Eq.
(\ref{masterequation11}) is the thermal state. From Eqs.
(\ref{averphoton0}) and (\ref{avephot1}) the thermal average photon
numbers $\langle \hat n\rangle_{th} =\frac{a _{11} +
a_{22}+\texttt{Re}[a_{23}]}{a _{44} - a _{11}}$ for
${\hat{\rho}_{AB}}$ corresponds to the inverse temperatures of
cavity field
\begin{eqnarray}\label{cavitytemp1}
{\beta}_{coh}=-\frac{1}{\omega}\ln\frac{\langle \hat
n\rangle_{th}}{1+\langle \hat n\rangle_{th}
}=-\frac{1}{\omega}\ln\frac{p_{e}+\texttt{Re}[a_{23}]}{p_{g}+\texttt{Re}[a_{23}]},
\end{eqnarray}
and $\langle \hat n\rangle_{th} ^{(m)}=\frac{a _{11} + a_{22}}{a
_{44} - a _{11}}$ for $\hat{\rho}_{AB}^{(m)}$ (m=1,2) with
\begin{eqnarray}\label{cavitytemp2}
{\beta}_{non}^{(m)}=-\frac{1}{\omega}\ln\frac{\langle \hat
n\rangle_{th} ^{(m)}}{1+\langle \hat n\rangle_{th}
^{(m)}}=-\frac{1}{\omega}\ln\frac{p_{e}}{p_{g}}.
\end{eqnarray}
From Eqs. (\ref{bathtemp}-\ref{cavitytemp2}) we can see that for the
noncoherent reservoir with ${\hat{\rho}_{AB}^{(m)} }$ the cavity
field is thermalized and reaches the same temperature as the
reservoir, ${\beta}_{non}^{(m)}=\beta_{\textrm{eff}}$, but for the
coherent reservoir with ${\hat{\rho}_{AB}}$ the temperature of
cavity field after thermalization does not coincide with the
reservoir's any more due to the reservoir's coherence, i.e.,
${\beta}_{coh}\neq \beta_{\textrm{eff}}$.
%%~~~~~~~~~~~~~~~~~~~~~~~~~~~~~~~~~~~~~~~~~~~~~~~~~~~~~~~~~~~~~~~~~~~~~~~~~~~~~~~~~~~第 1 处修改~~~~~~~~~~~~~~~~~~~~~~~
%%~~~~~~~~~~~~~~~~~~~~~~~~~~~~~~~~~~~~~~~~~~~~~~~~~~~~~~~~~~~~~~~~~~~~~补充审稿意见1~与单原子库对比~~Scully model~~~~~~新添加部分~~~~~~~~
It is similar to the model of a single atom coherent reservoir in
Ref.~\cite{Scully} where Scully \emph{et al}. showed that the
detailed balance between photon absorption and emission could be
broken with the help of the coherent superposition of the two
(nearly degenerate) lower levels of a three-level atom. In our
model, the TLAP has four energy levels including a higher level with
double excitation $|ee\rangle$, a lower level with no excitation
$|gg\rangle$ and two degenerate intermediate levels respectively
corresponding to single excitation $|eg\rangle$ and $|ge\rangle$.
Here, we can also use the coherent superposition of the two
degenerate intermediate levels of the TLAP to break the detailed
balance. Meanwhile, the deviation away from thermal equilibrium is
completely determined by the real part of the coherence term (see
Eq. (\ref{cavitytemp1})), i.e.,
$\texttt{Re}[a_{23}]=|a_{23}|\cos\phi$ where we denote $\phi$ as the
relative phase between the two degenerate intermediate levels in the
TLAP.
%%~~~~~~~~~~~~~~~~~~~~~~~~~~~~~~~~~~~~~~~~~~~~~~~~~~~~~~~~~~~~~~~~~~~~~~~~~~~~~ 结束~~~~~~~~~~~~~~~~~~~~~~~~~~~~~~~~~~・~~~~~~~~~~~~~~~~~
 It is clear that the reservoir's
coherence in our model also plays an important role in the
thermalization of the cavity field.

In addition, from Eq. (\ref{avephot1}) it can be seen that the
cavity field possesses the same average photon number for
$\hat{\rho}_{AB}^{(1)}$ and $\hat{\rho}_{AB}^{(2)}$ which implies
that the classical correlation of the TLAP has no contribution to
the work capability of cavity field. Actually, even for the quantum
correlations including the QE and QD they are not always good
quantities to effectively reflect the contributions of the TLAPs to
the work capability of cavity field. Using the density matrix Eq.
(\ref{Xstate}), the quantum correlations between the two atoms in
each TLAP as measured by QE and QD can be calculated. We adopt
Wootter's concurrence \cite{Wootters} as entanglement measure. For
the density matrix Eq. (\ref{Xstate}), the concurrence is given by
\begin{equation}\label{concurence}
    C(\hat\rho_{AB})=2\max(0, |a_{23}|-\sqrt{a_{11}a_{44}},
    |a_{14}|-\sqrt{a_{22}a_{33}}).
\end{equation}
On the other hand, quantum discord captures all nonclassical
correlations between two two-level atoms~\cite{Ollivier}. For the X
state described by the density matrix Eq. (\ref{Xstate}), the
analytic expression of QD has been reported~\cite{Wang} and
expressed by
\begin{equation}\label{discord}
     {Q}(\hat\rho_{AB})=\min({Q_1}, {Q_2}),
\end{equation}
where
$Q_j=H(a_{11}+a_{33})+\sum_{i=1}^4\lambda_i\log_2\lambda_i+D_j$ with
$\lambda_i$ being the four eigenvalues of $\hat \rho_{AB}$,
$D_1(\tau)=H(\tau)$,
$D_2(\tau)=-\sum_{i=1}^4a_{ii}\log_2a_{ii}-H(a_{11}+a_{33})$ with
$\tau= (1+\sqrt{[1-2(a_{33}+a_{44})]^2+4(|a_{14}|+|a_{23}|)^2})/{2}$
and $H(\tau)=-\tau\log_2\tau-(1-\tau)\log_2(1-\tau)$.
%%~~~~~~~~~~~~~~~~~~~~~~~~~~~~~~~~~~~~~~~~~~~~~~~~~~~~~~~~~~~~~~~~~~~~~~~~~~~~~~~~~~~~~~~~~~~~~~~~~~
%% the specific example for the X-state show the zero discord and entanglement which is same as the state emilating the nondiagnol elements。
%%~~~~~~~~~~~~~~~~~~~~~~~~~~~~~~~~~~~~~~~~~~~~~~~~~~~~~~~~~~~~~~~~~~~~~~~~~~~~~~~~~~~~~~~~~~~~~~~~~~
For the convenience of discussion
%%~~~~~~~~~~~~~~~~~~~~~~~~~~~~~~~~~~~~~~~~~~~~~~~~~~~~~~~~~~~~~~~~~~~~~~~~~~~~~~~~~~~第 2 处修改~~~~~~~~~~~~~~~~~~~~~~~
%%~~~~~~~~~~~~~~~~~~~~~~~~~~~~~~~~~~~~~~~~~~~~~~~~~~~~~~~~~~~~~~~~~~~~~补充审稿意见~How to construct the X state ~~~~~~新添加部分~~~~~~~~
we give our construction of the general X-state. As we know, the space
of a general X-state is composed of two independent subspaces which are
spanned by the base vectors $\{|ee\rangle,|gg\rangle\}$ and
$\{|eg\rangle,|ge\rangle\}$, respectively. We can choose the
arbitrary state in each subspace to construct the X-state via the
direct sum as follows
%%~~~~~~~~~~~~~~~~~~~~~~~~~~~~~~~~~~~~~~~~~~~~~~~~~~~~~~~~~~~~~~~~~~~~~~~~~~~~~ 结束~~~~~~~~~~~~~~~~~~~~~~~~~~~~~~~~~~・~~~~~~~~~~~~~~~~~
\begin{widetext}
\begin{align}\label{re:Xstate}
\hat{\rho} _{AB}=&\cos^{2}\alpha \hat{\rho} _{1}'\oplus
\sin^{2}\alpha\hat{\rho} _{2}'
   = &\left( {\begin{array}{*{20}c}
   { \frac{1}{2} {\cos^{2}\alpha}(1 + {r}_1 \cos\theta_1)} & 0 &  0 & {\frac{1}{2}r_1 {\cos^{2}\alpha} \sin\theta_1  e^ {-i\varphi }}  \\
   0 & {\frac{1}{2} {\sin^{2}\alpha}(1 + {r}_2 \cos\theta_2)  } &  { \frac{1}{2}r_2 {\sin^{2}\alpha} \sin\theta_2  e^ {-i\phi } } & 0  \\
   0 & { \frac{1}{2}r_2{\sin^{2}\alpha} \sin\theta_2  e^ {i\phi }  } &  { \frac{1}{2} {\sin^{2}\alpha}(1 - {r}_2 \cos\theta_2)  } & 0  \\
   { \frac{1}{2}r_1 {\cos^{2}\alpha} \sin\theta_{1}e^ {i\varphi} } & 0 &  0 & {  \frac{1}{2} {\cos^{2}\alpha}(1 - r_1 \cos\theta_1)  }
 \end{array} } \right),
 \end{align}
 \end{widetext}
in which the density matrices $\hat{\rho}
_{1,2}'=\frac{1}{2}(\mathbb{I}_{1,2}+\vec{r}_{1,2}\cdot\vec{\sigma}_{1,2})$,
respectively, represent the arbitrary state in each subspace where
$\mathbb{I}_{1}(\mathbb{I}_{2})$ is a unit matrix in the state space
of $\hat{\rho} _{1}'(\hat{\rho} _{2}')$, $\vec{r}_{1,2}$ are the
Bloch sphere vectors, and $\vec{\sigma}_{1,2}$ are the pauli
matrices with $\hat\sigma_{1z}=|ee\rangle\langle ee|-
|gg\rangle\langle gg|$ and $\hat\sigma_{2z}=|eg\rangle\langle eg|-
|ge\rangle\langle ge|$. And we assume that the probability of the
TLAP in each subspace is respectively $\cos^{2}\alpha$
$(\sin^{2}\alpha)$. The parameters in the X-state of Eq.
(\ref{re:Xstate}) satisfy $\alpha\in[0,\pi/2]$,
$\theta_1,\theta_2\in[0,\pi]$, $\varphi, \phi\in[0,2\pi]$ and $r_1
,r_2\in[0,1]$ which guarantee the positivity, normalization and
trace preservation of $\hat{\rho} _{AB}$.
%%~~~~~~~~~~~~~~~~~~~~~~~~~~~~~~~~~~~~~~~~~~~~~~~~~~~~~~~~~~~~~~~~~~~~~~~~~~~~~~~~~~~第 3 处修改~~~~~~~~~~~~~~~~~~~~~~
%%~~~~~~~~~~~~~~~~~~~~~~~~~~~补充一些内容保持前后内容衔接连贯~~~~~~~~~~~~~~~~~~~~~~~~~~~~~~~~~~补充我们可以调节Xstate的参数找到一系列关联为零的态，作为例子... ~~~~~~新添加部分~~~~~~~~
By choosing proper parameters of the X-state in Eq.
(\ref{re:Xstate}) we can find some states which only possess
coherence but no quantum correlations. That is, these states have
nondiagonal elements but their concurrence (quantum entanglement)
and quantum discord are zero.
%%~~~~~~~~~~~~~~~~~~~~~~~~~~~~~~~~~~~~~~~~~~~~~~~~~~~~~~~~~~~~~~~~~~~~~~~~~~~~~ 结束~~~~~~~~~~~~~~~~~~~~~~~~~~~~~~~~~~・~~~~~~~~~~~~~~~~~
 As an example, we choose
the parameters: $ r_1 =r_2 =2/3$, $\alpha= \sqrt{\pi}/4 $, $\theta_1
= \theta_2 = 11 \pi/20$, $\varphi = \pi/3$, $ \phi_2 = 0$, and the
X-state becomes
\begin{widetext}
\begin{align}\label{exampletho}
\hat{\rho} _{1}   = \left( {\begin{array}{*{20}c}
   { 0.223928  } & 0 &  0 & { 0.0823074-0.142561 i }  \\
   0 & { 0.223928  } &  {0.164615 } & 0  \\
   0 & {0.164615 } &  { 0.276072  } & 0  \\
   { 0.0823074+0.142561 i } & 0 &  0 & { 0.276072  }
 \end{array} } \right).
 \end{align}
\end{widetext}
From Eqs. (\ref{concurence}) and (\ref{discord}) we can obtain that
$C(\hat{\rho} _{1})=Q(\hat{\rho} _{1})=0$. It is noted that although
the state $\hat{\rho}^{\texttt{dia}} _{1}$, only preserving the
diagonal elements of density matrix in Eq. (\ref{exampletho}), has
no quantum correlations like $\hat{\rho}_{1}$ the average photon
numbers of cavity field in the two cases are different. This mean
that quantum coherence can be the work resource even in the absence
of quantum correlations.

%This means
%that the quantum correlations do not effectively reflect the
%reservoir's contribution to the work capability of cavity field.

%%~~~~~~~~~~~~~~~~~~~~~~~~~~~~~~~~~~~~~~~~~~~~~~~~~~~~~~~~~~~~~~~~~~~~~~~~~~~~~~~~~~~~~~~~~~~~~~~~~~
%% 重点介绍 x态下的平均光子数和另外两种的对比，说明相干干涉在其中的作用。单库热循环的设计
%%~~~~~~~~~~~~~~~~~~~~~~~~~~~~~~~~~~~~~~~~~~~~~~~~~~~~~~~~~~~~~~~~~~~~~~~~~~~~~~~~~~~~~~~~~~~~~~~~~~
 Meanwhile, although the states of the TLAPs, $\hat{\rho}_{AB}$ and
 $\hat{\rho}_{AB}^{(m)}$ (m=1,2) have the same reduced density matrices
 as mentioned before, they lead to different average photon numbers of
 cavity field $\langle \hat{n}\rangle_{ss}$
 and $\langle \hat{n}\rangle_{ss}^{(m)}$ given in Eqs. (\ref{averphoton0}) and (\ref{avephot1}), respectively.
 It is noted that from Eqs. (\ref{averphoton0}) and (\ref{avephot1})
$\langle \hat{n}\rangle_{ss}\geq\langle
 \hat{n}\rangle_{ss}^{(m)}$ for $a_{23}+a_{32}\geq 0$; $\langle \hat{n}\rangle_{ss}<\langle \hat{n}\rangle_{ss}^{(m)}$
for  $a_{23}+a_{32}< 0$. So, it means that due to the coherence of
the TLAPs not only the constructive quantum interference but also
the destructive quantum interference can be induced and cause the
average photon number of cavity field away from $\langle
\hat{n}\rangle_{ss}^{(m)}$ corresponding to $\hat{\rho}_{AB}^{(m)}$
(m=1,2) without any coherence. This is an interesting and meaningful
thing, and by using it we can easily perform a thermodynamic cycle
with a single nonequilibrium reservoir only by controlling an
external parameter, i.e., the relative phase. Furthermore, in terms
of the definitions of quantum correlations in Eqs.
(\ref{concurence}) and (\ref{discord}), QE and QD are only related
to the amplitude of the nondiagonal terms in the X-state and
independent of the relative phase. It demonstrates that quantum
correlations could not always reflect the effects of reservoir on
the system's work capability completely, and the relative phase of
the TLAP also plays an important role.

%It demonstrates again that reservoir's quantum correlations do not
%always serve as the resource of system's work capability indeed.

%%~~~~~~~~~~~~~~~~~~~~~~~~~~~~~~~~~~~~~~~~~~~~~~~~~~~~~~~~~~~~~~~~~~~~~~~~~~~~~~~~~~~~~~~~~~~~~~~~~~~~第 4 处修改~~~~~~~~~~~~~~~~~~~~~~
%%~~~~~~~~~~~~~~~~~~~~~~fast passing and no a14 ~~~~~here, it is proper to make comparision with lutz's~~~~~~~~~~~~~~~~~~~添加Lutz等人和我们的结果不矛盾，如何展示关联不是好的量~~~~~~新添加部分~~~~~

Moreover, we also notice that Dillenschneider \emph{et al}. in
Ref.~\cite{Lutz} also considered the same model where a series of
TLAPs pass through a cavity. They showed that the quantum
correlations act as the resource of system's work capability which
seems to be contradictory with our results (the reservoir's
coherence acting as the resource rather than the quantum
correlations). In fact, it is not that case. This can be explained
as follows. The state of TLAP considered in Ref.~\cite{Lutz} is a
thermal entangled state which belongs to a very special X-state with
$a_{22}= a_{33}$, $a_{14}= a_{41}=0$ and $a_{23}= a_{32}=-|a_{23}|$.
For this state we can obtain the average photon number of the cavity
field in the final steady state as
\begin{equation}\label{thermalphoto2}
   \langle \hat n_1\rangle_{ss} =\frac{a _{11} +a_{22}-|a_{23}|}{a _{44} - a _{11}}= \langle \hat n\rangle_{det}+\Delta,
\end{equation}
where $ \langle \hat n\rangle_{det}=\frac{a _{11} +a_{22}}{a _{44} -
a _{11}}$ represents the average photon number for the system being
in the detailed balance, and $\Delta=-\frac{|a_{23}|}{a _{44} - a
_{11}}$ denotes the deviation from the detailed balance. Since
$|a_{23}|$ can effectively reflect the quantum correlations of the
thermal entangled X-state the deviation term, $\Delta$, can not only
be understood as the thermodynamic effects of quantum coherence on
cavity field but also the contribution of quantum correlations. In
order to demonstrate that quantum correlations are not good physical
quantities to effectively reflect the thermodynamic effects of
nonequilibrium reservoir, we could implement a quantum phase gate
operation on one of the atoms in the TLAP to make the coherence term
$a_{23} $ $(a_{32})$ of the thermal entangled state add a relative
phase factor, i.e., the transition $a_{23}$ $(a_{32})\rightarrow
e^{-i\phi}a_{23}$ $(e^{i\phi}a_{32})=-e^{-i\phi}|a_{23}|$
$(-e^{i\phi}|a_{32}|)$, $\phi\in[0,\pi]$. It is noted that this
operation only adds a relative phase and does not change the quantum
correlations (concurrence and QD) of the TLAP. For the state after
the operation the average photon number of the cavity field in the
steady state can be expressed as
\begin{equation}\label{thermalphoto3}
   \langle \hat n_2\rangle_{ss} = \langle \hat n\rangle_{det}+\Delta
   \cos \phi.
\end{equation}
From Eq. (\ref{thermalphoto3}) it is clear that, for the average
photon number of the cavity field, the deviation from the detailed
balance not only depends on the quantum correlations but also on the
relative phase, and via modifying the relative phase the instructive
and the destructive interferences could be introduced. Comparing
Eqs. (\ref{thermalphoto2}) and (\ref{thermalphoto3}) we can see that
even though the TLAPs have the same quantum correlations they could
correspond to different average photon number, i.e., $\langle \hat
n_1\rangle_{ss}\neq\langle \hat n_2\rangle_{ss}$ for $\phi\neq 0$.
This demonstrates that, in general, the reservoir's quantum
correlations could not completely reflect the influence of a
nonequilibrium reservoir on the work capability of the cavity field,
and the relative phase also plays an important role. And by
modifying the relative phase the deviation from the detailed balance
for the average photon number may be positive, negative or zero.

%%~~~~~~~~~~~~~~~~~~~~~~~~~~~~~~~~~~~~~~~~~~~~~~~~~~~~~~~~~~~~~~~~~~~~~~~~~~~~~ 结束~~~~~~~~~~~~~~~~~~~~~~~~~~~~~~~~~~・~~~~~~~~~~~~~~~~~

%%~~~~~~~~~~~~~~~~~~~~~~~~~~~~~~~~~~~~~~~~~~~~~~~~~~~~~~~~~~~~~~~~~~~~~~~~~~~~~~~~~~~~~~~~~~~~~~~~~~
%%从相干导致的不同光子数，引出a14，a41，过渡到非近似下的处理结果
%%~~~~~~~~~~~~~~~~~~~~~~~~~~~~~~~~~~~~~~~~~~~~~~~~~~~~~~~~~~~~~~~~~~~~~~~~~~~~~~~~~~~~~~~~~~~~~~~~~~

From Eq. (\ref{averphoton0}) we can see that the coherence elements
$a_{14}$ and $a_{41}$ do not like $a_{23}$ and $a_{32}$ contribute
to the average photon number $\langle \hat{n}\rangle_{ss}$. They
seem to have no effect on the average photon number. In fact, it is
not that case. This can be explained that $a_{14}$ ($a_{41}$)
corresponds to the process of the double excitation during the
evolution of the cavity field, and appears in the higher order terms
than the second order term $\xi^{2}$ of the average photon number.
In the limit $\xi\rightarrow 0$, these high order terms can be
omitted, however, they can not be ignored any more in case 2 with
finite $\xi$. In case 2 the effects of coherence $a_{14}$ $(a_{41})$
on the dynamics of the cavity field will be fully demonstrated.
%%%%%%%%%%%%%%%%%%%%%%%%%%%%%%%%%%%%%%%%%%%%%%%%%%%%%%%%%%%%%%%%%%%%
\begin{figure} \includegraphics[width=8 cm]{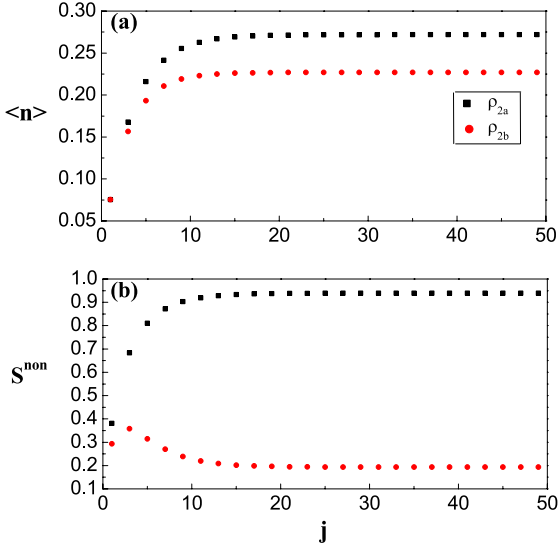}
 \caption{(Color online). (a) The variations of the average photon number, $\langle\hat n\rangle$, and (b)
 the variations of the entropy,
 $S^{non}$, of the cavity field with passing times, $j$, for state $\hat\rho_{2a}$ (black squares)
and $\hat\rho_{2b}$ (red dots).} \label{fig2a}
\end{figure}

%%%%%%%%%%%%%%%%%%%%%%%%%%%%%%%%%%%%%%%%%%%%%%%%%%%%%%%%%%%%%%%%%%%%

From Eq. (\ref{evolution22}) we can see that the expression of the
density matrix of the cavity after the $(j+1)$th passing of the
TLAPs is very complicated, and the nondiagonal elements appear in the
density matrix which is very different from that in case 1. Thus, it
is very difficult to obtain general analytical expressions of the average
photon number and the entropy of the cavity field during the evolution.
Next, we will make use of numerical calculations to explore the
dynamics of the cavity field. In order to demonstrate the effects of
coherence terms $a_{14}$ and $a_{41}$ in the X-state given in Eq.
(\ref{Xstate}) we consider the TLAPs being respectively prepared in
the following two states
\begin{align}\label{tho2b}
\hat{\rho} _{2a}   = \left( {\begin{array}{*{20}c}
   { 0.142864  } & 0 &  0 & {0}  \\
   0 & {0.0122355 } &  {0.0012236 } & 0  \\
   0 & {0.0012236 } &  { 0.0122355 } & 0  \\
   { 0 } & 0 &  0 & { 0.832665  }
 \end{array} } \right)
 \end{align}
and
\begin{align}\label{tho2a}
\hat{\rho} _{2b}   = \left( {\begin{array}{*{20}c}
   { 0.142864  } & 0 &  0 & { 0.344901 }  \\
   0 & {0.0122355 } &  {0.0012236 } & 0  \\
   0 & {0.0012236 } &  { 0.0122355 } & 0  \\
   { 0.344901 } & 0 &  0 & { 0.832665  }
 \end{array} } \right),
 \end{align}
where from Eq. (\ref{re:Xstate}) we choose the parameters $\{ r_1 =0.7071,r_2
=0.1,\alpha=\pi/20 ,\theta_1 =\pi, \theta_2 = \pi/2,\varphi = \phi =
0\}$ for $\hat\rho_{2a}$ and $\{r_1 =1,r_2
=0.1,\alpha=\pi/20 ,\theta_1 =3\pi/4, \theta_2 = \pi/2,\varphi =
\phi = 0\}$ for $\hat{\rho}_{2b}$. It is noted that $a_{14}= 0$ in
$\hat\rho_{2a}$ and $a_{14}= 0.344901$ in $\hat\rho_{2b}$ and the
other density matrix elements are the same. Hereafter, we set
$\xi=0.5$. In terms of Eq. (\ref{evolution22}) we plot the
variations of the average photon number, $\langle\hat {n}\rangle$,
and the entropy, $S^{non}$, of the cavity field with passing times, $j$,
for $\hat\rho_{2a}$ and $\hat\rho_{2b}$ in Fig. 2. Throughout this
paper, the entropy of the cavity field is defined by von Neumann entropy
$S(\rho)=-k_B\texttt{Tr}(\rho\log_2\rho)$, and for simplicity, we
set $k_B=1$. From Fig. 2 it can be seen that the nondiagonal element
$a_{14}$ ($a_{41}$) has an obvious effects on the average photon
number shown in Fig. 2(a) and the entropy of the cavity field in Fig.
2(b), and the values of $\langle\hat {n}\rangle$ and $S^{non}$, for
$\hat\rho_{2a}$ are always larger than those for $\hat\rho_{2b}$.
Moreover, we also see that the changes of the entropy and the
average photon number of cavity field are obvious for the first few
passings and for $j>10$ the entropy and the average photon number
gradually approach their steady values, respectively.

%%~~~~~~~~~~~~~~~~~~~~~~~~~~~~~~~~~~~3.24 新添加段落~~~~~~~~~~~~~~~~~~~~~~~~~~~~~~~~~~~~~~~~~~~~~~~~~~~~~~~~~~~~~~~~~~~~~~~~~~~~~~~~~~~~~~~~~~~~~~~~~~~~~~~~~~~~~~~~~~
On the other hand, $\xi$ could also affect the dynamic behavior of
the cavity field, such as the average photon number and the entropy.
Different from case 1 $\xi\rightarrow 0$ $(\xi=g\tau)$  the cavity
field could exhibit very different and complicated dynamic behaviors
in case 2 with  finite $\xi$ even though we choose the same
nonequilibrium reservoir. The finite $\xi$ could be understood as a
strong coupling or a long interaction time interval between the TLAP
and the cavity field, and in this case the double excitation process
corresponding to the coherent term $a_{14}$ $(a_{41})$ is involved
in the dynamic evolution of the cavity field in case 2, while for
$\xi\rightarrow 0$ in case 1 this can not occur. According to the
unitary operator $\hat{U}(\tau)$ in Eq. (\ref{unitary5}) we can see
that each element of $\hat{U}(\tau)$ in case 2 is a nonlinear
function of $\xi$ that could lead to a very complicated evolution of
the cavity field, $\hat{\rho}(t_j+\tau)$ in Eq. (\ref{evolution22}).
From the density matrix $\hat{\rho}(t_j+\tau)$ it can be seen that
the single excitation process corresponding to the coherent term
$a_{23}$ $(a_{32})$, the double excitation process with coherence
$a_{14}$ $(a_{41})$ and the parameter $\xi$ are all involved in the
cavity evolution in a very complicated way. Naturally, from the
definitions of the average photon number and the entropy of the
cavity field, we can infer that during the cavity evolution they are
also closely related to the single excitation, double excitation
processes, and the parameter $\xi$. Since the expressions of the
average photon number and the entropy of the cavity field are very
complicated we can only demonstrate how the parameter $\xi$ and
reservoir's coherence $a_{23}$ and $a_{14}$ influence the dynamics
of the cavity by numerical calculations. Only as an example, we plot
Fig. 2 to demonstrate the thermodynamic effects of the coherent term
$a_{14}$ (or the double excitation process of the TLAP) on the
dynamics of the cavity field for specific states of the TLAP
$\hat{\rho}_{2a}$ and $\hat{\rho}_{2b}$, and $\xi$ ($\xi=0.5$). From
a lot of numerical calculations we find that the dynamic behaviors
of the cavity field are determined by $a_{23}$, $a_{14}$, and $\xi$
altogether. For example, suppose that the state of the TLAP is
$\hat{\rho}_{AB}$ when we keep $\hat{\rho}_{AB}$ fixed and only
change the value of $\xi$, or keep $\xi$ fixed and only change the
values of the non-diagonal elements of $\hat{\rho}_{AB}$ the average
photon number and the entropy of the cavity field usually could
exhibit different variation curves with the passing times $j$, i.e.,
the entropy and the average photon number of the cavity field can
exhibit monotonous behaviors for some specific $\xi$ and specific
X-states of the TLAP, and non-monotonous behaviors for the others.
For example, if we choose the state of the TLAP as $\hat{\rho}_{2b}$
given in Eq. (\ref{tho2a}) and $\xi=0.2$ instead of $\xi=0.5$ in
Fig. 2 the entropy of the cavity will be monotonously increasing
with $j$ which is very different from the non-monotonous behavior as
shown in Fig. 2(b). Similarly, the monotonous increasing behavior of
the entropy can also appear only by choosing the proper state of the
TLAP and keeping $\xi$ unchanged such as $\xi=0.5$ and changing
$\hat{\rho}_{2b}$ in Eq. (\ref{tho2a}) into the state
$\hat{\tilde{\rho}}_{2b}$ with $\{ r_1 =0.7071,r_2
=0.1,\alpha=\pi/4,\theta_1 =3\pi/4, \theta_2 = \pi/2,\varphi = \phi
= 0\}$ in Eq. (\ref{re:Xstate}). In Fig. 2(a) although the average
photon number $\langle n \rangle$ exhibits a monotonous increasing behavior for
$\hat{\rho}_{2b}$ and $\xi=0.5$ the curve of $\langle n\rangle$ can also exhibit
the non-monotonous behavior for other value of  $\xi$ and other
state of the TLAP, such as $\xi=1.2$ and $\hat{\rho}_{AB}$ in Eq.
(\ref{re:Xstate}) with $\{ r_1 =1,r_2 =0.5,\alpha=\pi/20,\theta_1
=3\pi/4, \theta_2 = \pi/2,\varphi = \phi = 0\}$. In addition, from
numerical calculations we also find that if $a_{14}=0$ both the
average photon number and the entropy of the cavity always
monotonously increase with $j$ which indicates that the coherent
term $a_{14}$ may effectively influence the dynamic behaviors of the
cavity field in case 2 with finite $\xi$. This means that the
non-monotonous behavior of the cavity field might be caused by the
double excitation process corresponding to the coherent term
$a_{14}$ for finite $\xi$.

Besides, for a cavity field coupled to a non-equilibrium atomic
reservoir with effective temperature well defined by the two-level
atom if the detailed balance is not broken the temperature of the
cavity field being in a steady state will be the same as the
effective temperature of the reservoir, and the cavity has the
average photon number defined by the effective temperature of the
reservoir. For the cavity field with the TLAPs being initially in
the X-state $\hat{\rho}_{AB}$ in Eq. (\ref{Xstate}) if the detailed
balance is not broken the average photon number of cavity field will
only depend on the diagonal elements of $\hat{\rho}_{AB}$. In
another word, if other parameters such as the coherence of the
non-equilibrium reservoir, i.e., the non-diagonal elements of
$\hat{\rho}_{AB}$, or the parameter $\xi$ enter into the average
photon number of the cavity field the detailed balance in general
could not be reached. From the density matrix of the cavity
evolution $\hat{\rho}(t_j+\tau)$ in Eq. (\ref{evolution22}) we can
see that the density matrix of the cavity field in case 2 always has
nonzero nondiagonal elements which means that the cavity field
always stays in a nonequilibrium state even if it reaches a steady
state. In addition, we also notice that in case 2 all the coherence
of the TLAP $a_{23}$ and $a_{14}$ have been involved in the diagonal
elements of $\hat{\rho}(t_j+\tau)$ which means that the average
photon number during the cavity evolution will always carry the
reservoir's coherence information. Thus, the detailed balance in
case 2 is usually broken by the reservoir's coherence $a_{23}$ and
$a_{14}$ cooperatively, which is different from case 1 where only
the reservoir's coherence $a_{23}$ corresponding to the single
excitation process breaks the detailed balance.

%%~~~~~~~~~~~~~~~~~~~~~~~~~~~~~~~~~~~~~~~~~~~~~~~~~~~~~~~~~~~~~~~~~~~~~~~~~~~~~ 结束~~~~~~~~~~~~~~~~~~~~~~~~~~~~~~~~~~・~~~~~~~~~~~~~~~~~

%%%%%%%%%%%%%%%%%%%%%%%%%%%%%%%%%%%%%%%%%%%%%%%%%%%%%%%%%%%%%%%%%%%%
\begin{figure} \includegraphics[width=8 cm]{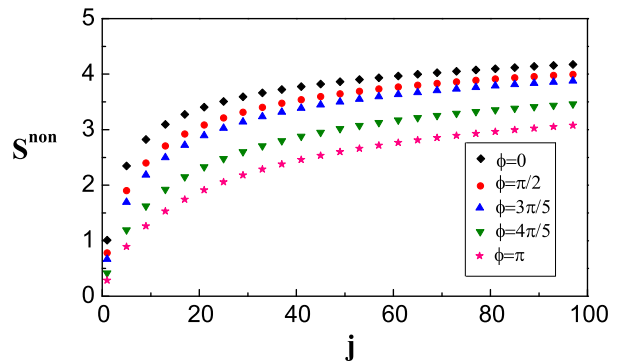}
 \caption{(Color online). The variations of the entropy of the cavity field, $S^{non}$, with passing times of the TLAPs, $j$,
 for different relative phases, $\phi=\{0, \pi/2, 3\pi/5, 4\pi/5,
 \pi\}$, and the other parameters of the state in Eq. (\ref{re:Xstate}) are $r_1 =0.2$, $r_2
=1$, $\alpha=\pi/3$, $\theta_1 =3\pi/4$, $\theta_2 = \pi/2$,
$\varphi = 0$.} \label{fig3a}
\end{figure}
%%%%%%%%%%%%%%%%%%%%%%%%%%%%%%%%%%%%%%%%%%%%%%%%%%%%%%%%%%%%%%%%%%%%%%%%%%

%%~~~~~~~~~~~~~~~~~~~~~~~~~~~~~~~~~~~~~~~~~~~~~~~~~~~~~~~~~~~~~~~~~~~~~~~~~~~~~~~~~~~~~~~~~~~~~~~~~~
%% the effects of coherence relative pahse 固定\varphi=0； discus different \phi 对 n， S影响。
%% 22 上午
%%~~~~~~~~~~~~~~~~~~~~~~~~~~~~~~~~~~~~~~~~~~~~~~~~~~~~~~~~~~~~~~~~~~~~~~~~~~~~~~~~~~~~~~~~~~~~~~~~~~

Next, let us consider the influence of the relative phase in the coherent
term $a_{23}$ of the TLAP on the work capability of the cavity field. As
mentioned before, although the quantum correlations, QE and QD, have
nothing to do with the relative phase in $a_{23}$ $(a_{32})$, it is
very important to determine the constructive or the destructive
interference in the work capability of cavity field as shown in Eq.
(\ref{averphoton0}) in the limit $\xi\rightarrow 0$. For case 2 with
finite $\xi$ it is difficult to obtain explicit expressions of the
average photon number and the entropy of the cavity, but we can
demonstrate the effects of the relative phase on the entropy of the cavity
field via numerical calculations. In terms of Eqs.
(\ref{evolution22}) and (\ref{re:Xstate}) we plot the variations of
the entropy of the cavity field, $S^{non}$, with passing times, $j$, for
different relative phases, $\phi=\{0, \pi/2, 3\pi/5,4\pi/5, \pi\}$
in Fig. 3 where the other parameters are: $r_1 =0.2$, $r_2 =1$,
$\alpha=\pi/3$, $\theta_1 =3\pi/4$, $\theta_2 = \pi/2$, $\varphi =
0$. From Fig. 3 we can see that for fixed passing time, $j$, the
entropy of cavity field decreases with the relative phase $\phi$ for
$\phi\in[0,\pi]$. In fact, when the relative phase ranges from 0 to
$2\pi$ we can find that for fixed passing time, $j$, the entropy of
cavity field $S^{non}$ is symmetric about $\phi=\pi$, i.e.,
$S^{non}(\pi-\phi)=S^{non}(\pi+\phi)$ with $\phi\in[0,\pi]$.
Moreover, Fig. 3 also shows that for an arbitrary $\phi$ the entropy
of the cavity field always increases with the passing times, $j$, and
its increment for the neighboring twice passings decreases which
means that the entropy, $S^{non}$, is convergent as expected.

%~~~~~~~~~~~~~~~~~~~~~~~~~前面分析的总结部分~~~~~~~~~~~~~~~~~~~~~~~~~~~~~~~~~~~~~~~~~~~~~~~~~~~~~~~~~~~~~~~~~~~~
Based on the above analysis, we can see that no matter whether there
exist quantum correlations or not the reservoir's coherence could
have an effect on the work capability of the cavity. Moreover, it
has been shown that although the relative phase is independent of
the quantum correlations it has an important effect on the dynamics
of cavity field. The constructive and destructive interferences
could be induced to change the thermodynamic features of cavity
field, such as the entropy and the average photon number of cavity,
via controlling the relative phase. It is obvious that the
reservoir's coherence plays a central role in system's work
capability in our model, and it could be taken as an effective
source of system's work capability even in the absence of quantum
correlations. This is our major result in this paper.

%%~~~~~~~~~~~~~~~~~~~~~~~~~~~~~~~~~~~~~~~~~~~~~~~~~~~~~~~~~~~~~~~~~~~~~~~~~~~~~~~~~~~~~~~~~~~~~~~~~~
%% 关于potential work capabilities 阐释 n 与 S相比， S更本质。
%%~~~~~~~~~~~~~~~~~~~~~~~~~~~~~~~~~~~~~~~~~~~~~~~~~~~~~~~~~~~~~~~~~~~~~~~~~~~~~~~~~~~~~~~~~~~~~~~~~~
It is emphasized that the cavity is always in a nonequilibrium
state with the TLAPs passing. As mentioned before, only for certain
conditions $\hat{\rho}_{A}=\hat{\rho}_{B}$ and
$a_{11}+a_{22}<a_{33}+a_{44}$ the steady state of the cavity in case 1
becomes a thermal equilibrium state. In general, the work capability
refers to the entropy of the cavity field not to the average photon number.
Roughly speaking, when the density matrix of the cavity keeps in
diagonal distribution the average photon number of cavity has
similar behavior as the entropy of cavity (i.e., when the average
photon number of cavity increases the corresponding entropy of
cavity will also increase), and it can also be used to describe the
work capability of cavity. Strictly speaking, it is not precise to
use the average photon number describing the potential work
capability of cavity because it can not effectively determine the
actual degree of work capability of cavity in a nonequilibrium
state.

%%~~~~~~~~~~~~~~~~~~~~~~~~~~~~~~~~~~~~~~~~~~~~~~~~~~~~~~~~~~~~~~~~~~~~~~~~~~~~~~~~~~~第 4 处修改~~~~~~~~~~~~~~~~~~~~~~~
%%~~~~~~~~~~~~~~~~~~~~~~~~~~~~~~~~~~~~~~~~~~~~~~~~~~~~~~~~~~~~~~~~被修改部分~~~~~~~~~~~~~~~~~~~~~~~~~~~~~~~~~~・~~~~~~~~~~~~~~

Especially, when the density matrix of the cavity has nondiagonal
elements the change tendency of the average photon number of the cavity
field can not always coincide with that of the entropy of the cavity field
as shown in Fig. 2.
%%~~~~~~~~~~~~~~~~~~~~~~~~~~~~~~~~~~~~~~~~~~~~~~~~~~~~~~~~~~~~~~~~~~~~~~~~~~~~~~修改后~~~~~~~~~~~~~~~~~~~~~~~~~~~~~~~~~~・~~~~~~~~~~~~~~
Only for the cavity being in a thermal state although the average
photon number denoted as $\bar{n}$ and the entropy denoted as $S$
are different physical quantities they may have similar behavior which can be
explained as follows. As we know that the
probabilities $P_m$ for the cavity field being in a thermal state with $m=0,1,2,...$ photons can be expressed as
\begin{equation}\label{photonproba}
P_m=\frac{\bar{n}^m}{(\bar{n}+1)^{m+1}},
\end{equation}
which satisfy the normalization condition $\sum_m P_m=1 $. Inserting
Eq. (\ref{photonproba}) into the entropy expression of the cavity field
$S=-\sum_m P_m\log_2P_m$ and after some calculations one obtains
\begin{equation}\label{thermentropy1}
S=(\bar{n}+1)\log_2(\bar{n}+1)- \bar{n} \log_2\bar{n}.
\end{equation}
From Eq. (\ref{thermentropy1}) it is easy to verify that
${dS}/{d\bar{n}}>0$ which means that the entropy of the cavity field
in a thermal state, $S$, is a monotonous increasing function of
the average photon number, $\bar{n}$. So, both the average photon
number and the entropy of the cavity field in the thermal state can
be used to describe the work capability of cavity.
%%~~~~~~~~~~~~~~~~~~~~~~~~~~~~~~~~~~~~~~~~~~~~~~~~~~~~~~~~~~~~~~~~~~~~~~~~~~~~~~~ 结束~~~~~~~~~~~~~~~~~~~~~~~~~~~~~~~~~~・~~~~~~~~~~~~~~~~~
%%%%%%%%%%%%%%%%%%%%%%%%%%%%%%%%%%%%%%%%%%%%%%%%%%%%%%%%%%%%%%%%%%%%
\begin{figure} \includegraphics[width=8 cm]{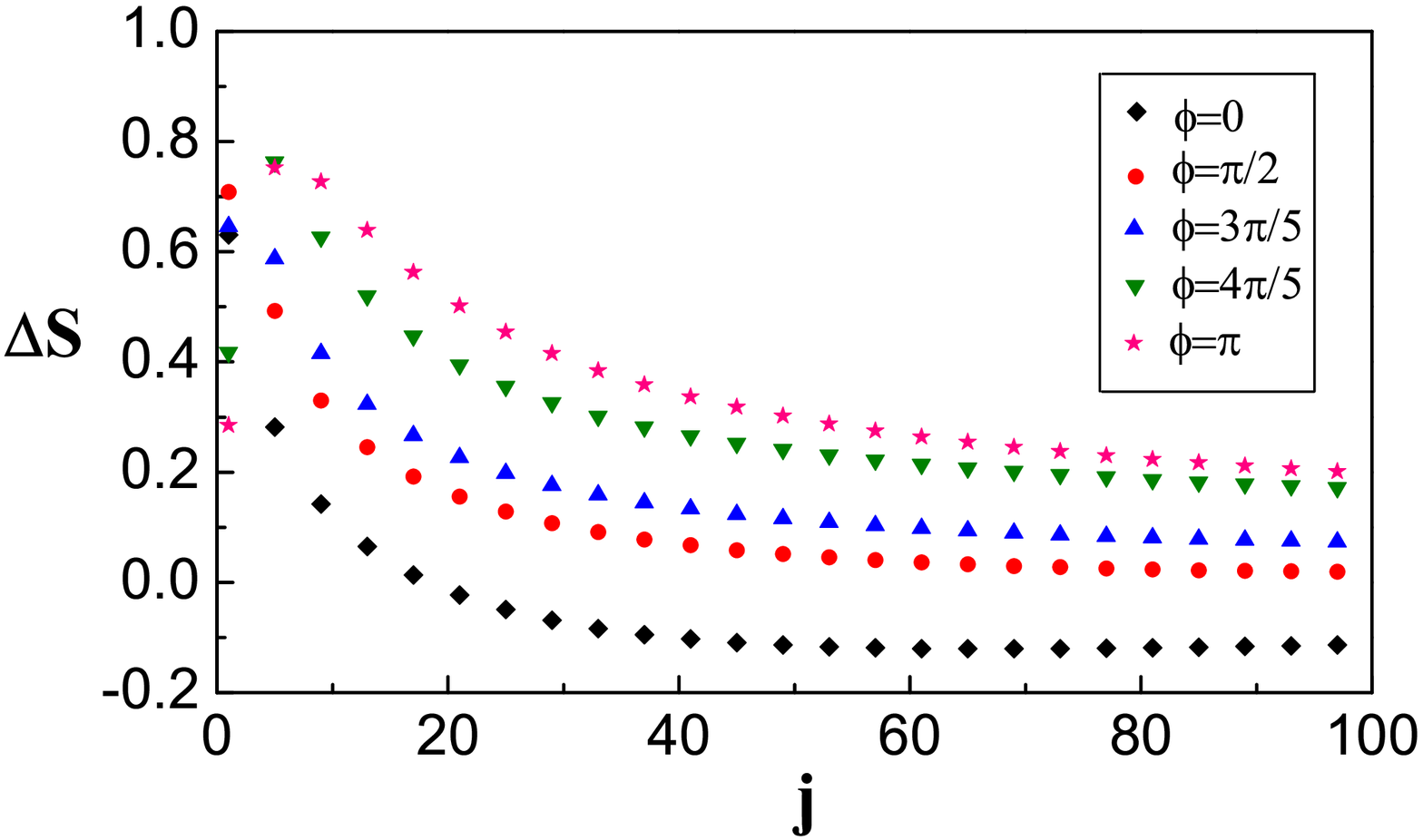}
 \caption{(Color online). The variations of the entropy difference of cavity field, $\triangle S$, with passing times of the TLAPs, $j$,
 for different relative phases, $\phi=\{0, \pi/2, 3\pi/5, 4\pi/5,
 \pi\}$, and the other parameters of the state in Eq. (\ref{re:Xstate}) are $r_1 =0.2$, $r_2
=1$, $\alpha=\pi/3$, $\theta_1 =3\pi/4$, $\theta_2 = \pi/2$,
$\varphi = 0$.} \label{fig3b}
\end{figure}
%%%%%%%%%%%%%%%%%%%%%%%%%%%%%%%%%%%%%%%%%%%%%%%%%%%%%%%%%%%%%%%%%%%%%%%%%%
Thus, we argue that in general the proper physical quantity to
measure the work capability of cavity is the entropy of cavity
rather than the average photon number. In order to make this point
more clearer we perform a thermalization on the cavity when the
$j$th TLAP passes through the cavity, and keep the average photon
number of the cavity unchanged, i.e., the energy of the cavity
remains constant during the thermalization. Denote the average
photon number and the density matrix of the cavity, after the $j$th
passing, as ${\bar{n}_j}$ and ${\hat{\rho}_j}$ satisfying Eq.
(\ref{evolution22}) and the state after thermalization as
${\hat{\rho}_j^{th}}$. Then, the thermal state of cavity,
${\hat{\rho}_j^{th}}$, can be expressed as
\begin{equation}\label{entropy}
    {\hat{\rho}_j^{th}}=Z^{-1} \sum_{n'=0}^{\infty}e^{-
    n'/\bar{n}_j}|n{'}\rangle\langle n{'}|,
\end{equation}
where $Z=1-e^{-1/\bar{n}_j}$ and
$\bar{n}_j=\sum_{n=0}^{2j+1}n\rho_{nn}$ with $n$ and $\rho_{nn}$
respectively being the energy level and the corresponding
probability distribution of the state $\hat{\rho}_j$. The entropy
difference of the cavity, ${\triangle S}$, between the states before and after thermalization is given by
\begin{equation}\label{entropydiff}
   {\triangle S}=S^{non}-S^{th},
\end{equation}
where $S^{non}$ and $S^{th}$ independently correspond to the von
Neumann entropy $S(\hat{\rho}_j)$ and $S(\hat{\rho}_j^{th})$. Here,
the entropy difference ${\triangle S}$ can effectively describe the
deviation away from the corresponding thermal equilibrium.
As an example, we choose the same parameters as that in Fig. 3, and
plot the variations of the entropy difference of the cavity with the passing
times $j$ for different relative phases $\phi=\{0, \pi/2, 3\pi/5,
4\pi/5,
 \pi\}$ in Fig. 4. From Fig. 4 we can see that during the evolution
 the entropy difference of the cavity may be larger or smaller than zero which
depends on the passing times of the TLAPs, $j$, and the relative
phase, $\phi$. That implies that the cavity might absorb heat from
the reservoir $(\triangle S>0)$ or release heat to the reservoir
$(\triangle S<0)$ during the thermalization process. Thus, it is the
entropy of cavity actually describing the potential work capability
of cavity not the average photon number except that the cavity
reaches thermal equilibrium, and in this case, although the average
photon number and the entropy are different physical quantities but
they have similar behavior as shown in Eq. (\ref{thermentropy1}).
This is consistent with the spirit demonstrated in another model
\cite{Esposito} most recently reported where the authors proposed a
entropic motor by exploiting entropy to fuel an engine and showed
that the generation of the entropic forces is surprisingly robust to
local changes in kinetic and topological parameters.

\section{summary and conclusions}{\label{Sec:4}

%%~~~~~~~~~~~~~~~~~~~~~~~~~~~~~~~~~~~~~~~~~~~~~~~~~~~~~~~~~~~~~~~~~~~~~~~~~~~~~~~~~~~~~~~~~~~~~~~~~~
%%~~~~~~~~~~~~~~~~~~~~~~~~~~~~~~~~~~~~~~~~~~~~~~~~~~~~~~~~~~~~~~~~~~~~~~~~~~~~~~~~~~~~~~~~~~~~~~~~~~

In conclusion, we have studied the dynamics of cavity with a
nonequilibrium reservoir consisting of a beam of identical TLAPs
initially preparing in the general X-state, and have derived a
quantum master equation. We have found that the coherence of a
nonequilibrium reservoir consisting of TLAPs in the X-state plays a
central role in the dynamics of cavity field. It has been shown that
not only the constructive interference but also the destructive
interference could be induced only by adjusting the relative phase
with which the quantum correlations have nothing to do. Using this
property a thermodynamic cycle with a single reservoir can be
implemented only via controlling one external parameter, the
relative phase. Meanwhile, we have also found that no matter whether
the quantum correlations exist or not the coherence of reservoir
could have contributions to the work capability of cavity. We, in
the present paper, have clearly demonstrated that quantum coherence
rather than quantum correlations can reflect the effects of
reservoir on the system's work capability effectively. In addition,
we have also shown that the proper physical quantity to measure the
potential work capability of cavity field is the entropy of cavity
field rather than the average photon number except that the cavity
arrives at thermal equilibrium, and in this case both of them can be
used to describe the work capability of cavity field. This work
might prompt further studies on how to use coherence as a
thermodynamic resource, such as the study of coherence in the heat
dissipation of atomic-scale junctions in most recent
experiments~\cite{Lee}. Finally, it is also interesting to extend
our present work to the Dicke model and some new results might be
revealed which will be considered in the future work.

\bigskip{}

%-------------------------ACKNOWLEDGMENTS--------------------------

\section*{ACKNOWLEDGMENTS}
This work is financially supported by National Science Foundation of
China (Grants Nos. 11274043, 11375025 and 61307041), the National
Science Foundation of Shandong Province, China (Grants No.
ZR2011FL009, ZR2013AQ013), the Science and Technology Project of
University in Shandong Province, China (Grant No. J12LJ01) and the
Youth Foundation of Shandong Institute of Business and Technology
(Grant No. 2013QN059).

%-------------------------ACKNOWLEDGMENTS--------------------------

\appendix
%-------------------------appendix1--------------------------------

\label{appendix1}

\section{The expressions of $f_{i}$}

The parameters $f_{i}$, ($i=1,2,3,...,11$), in Eq.
(\ref{evolution22}) are given by
\begin{equation}
\begin{aligned}
f_1=&a_{11}U_{11}(m)U_{11}^{}(n)+a_{44}U_{44}(m)U_{44}^{}(n)\\
+&(a_{22}+a_{33})[U_{22}(m)U_{22}^{}(n)+U_{23}(m)U_{23}^{}(n)]\\
+&(a_{23}+a_{32})[U_{22}(m)U_{23}^{}(n)+U_{23}(m)U_{22}^{}(n)],\\
f_2=&(a_{22}+a_{33}+a_{23}+a_{32})U_{12}(m+1)U_{12}^{*}(n+1)\\
+&2a_{44}U_{24}(m+1)U_{24}^{*}(n+1),\\
f_3=&(a_{22}+a_{33}+a_{23}+a_{32})U_{42}(m-1)U_{42}^{*}(n-1)\\
+&2a_{11}U_{21}(m-1)U_{21}^{*}(n-1),\\
f_4= &a_{11}U_{41}(m-2)U_{41}^{}(n-2),\\
 f_5=&a_{44}U_{14}(m+2)U_{14}^{}(n+2),\\
  f_6=&a_{14}U_{11}(m)U_{14}^{}(n+2),\\
   f_7=&a_{41}U_{14}(m+2)U_{11}^{}(n),\\
    f_8=&2a_{41}U_{24}(m+1)U_{42}^{*}(n-1),\\
f_9=&2a_{14}U_{24}(m-1)U_{42}^{*}(n+1),\\
 f_{10}=&2a_{41}U_{44}(m)U_{41}^{}(n-2),\\
  f_{11}=&2a_{14}U_{44}(m-2)U_{41}^{}(n),
\end{aligned}
\end{equation}
where $U_{ij}(x)$, $(i,j=1,2,3,4)$ ($x$ is the nonnegative integer )
are expressed as\\
\\
\\
\\
\\
\begin{equation}
\begin{aligned}
U_{11}(x)=&1+\frac{(x+1)[\cos g\tau\sqrt{2(2x+3)}-1]}{2x+3}\\
 U_{44}(x)= &1+\frac{x[\cos g\tau\sqrt{2(2x-1)}-1]}{2x-1} \\
U_{22}(x)= &U_{33}(x)=\frac{1}{2}{[\cos g\tau\sqrt{2(2x+1)}+1]}\\
U_{23}(x)= &U_{32}(x)=\frac{1}{2}{[\cos g\tau\sqrt{2(2x+1)}-1]}\\
U_{14}(x)=& \frac{\sqrt{x(x-1)}[\cos g\tau\sqrt{2(2x-1)}-1]}{2x-1}\\
U_{41}(x)=& \frac{\sqrt{(x+1)(x+2)}[\cos g\tau\sqrt{2(2x+3)}-1]}{2x+3}\\
U_{12}(x)=&=U_{13}(x)=-i\sqrt{x}\frac{\sin
g\tau\sqrt{2(2x+1)}}{\sqrt{2(2x+1)}}\\
U_{21}(x)=&=U_{31}(x)=-i\sqrt{x+1}\frac{\sin
g\tau\sqrt{2(2x+3)}}{\sqrt{2(2x+3)}}\\
U_{24}(x)=&=U_{34}(x)=-i\sqrt{x}\frac{\sin
g\tau\sqrt{2(2x-1)}}{\sqrt{2(2x-1)}}\\
U_{42}(x)=&=U_{43}(x)=-i\sqrt{x+1}\frac{\sin
g\tau\sqrt{2(2x+1)}}{\sqrt{2(2x+1)}}.\\
% \nonumber
\end{aligned}
\end{equation}

%\begin{widetext}
%\begin{equation}
%\begin{aligned}
%U_{11}(x)=&1+\frac{(x+1)[\cos g\tau\sqrt{2(2x+3)}-1]}{2x+3}
%~~~~ ~~U_{44}(x)= 1+\frac{x[\cos g\tau\sqrt{2(2x-1)}-1]}{2x-1}\\
%U_{22}(x)= &U_{33}(x)=\frac{1}{2}{[\cos g\tau\sqrt{2(2x+1)}+1]}
%~~~~~~U_{23}(x)= U_{32}(x)=\frac{1}{2}{[\cos g\tau\sqrt{2(2x+1)}-1]}\\
%U_{14}(x)=& \frac{\sqrt{x(x-1)}[\cos g\tau\sqrt{2(2x-1)}-1]}{2x-1}
%~~~~~~~U_{41}(x)=\frac{\sqrt{(x+1)(x+2)}[\cos g\tau\sqrt{2(2x+3)}-1]}{2x+3}\\
%U_{12}(x)=&=U_{13}(x)=-i\sqrt{x}\frac{\sin
%g\tau\sqrt{2(2x+1)}}{\sqrt{2(2x+1)}}
%~~~~U_{21}(x)=U_{31}(x)=-i\sqrt{x+1}\frac{\sin
%g\tau\sqrt{2(2x+3)}}{\sqrt{2(2x+3)}}\\
%U_{24}(x)=&=U_{34}(x)=-i\sqrt{x}\frac{\sin
%g\tau\sqrt{2(2x-1)}}{\sqrt{2(2x-1)}}
%~~~~U_{42}(x)=U_{43}(x)=-i\sqrt{x+1}\frac{\sin
%g\tau\sqrt{2(2x+1)}}{\sqrt{2(2x+1)}}.\\
%% \nonumber
%\end{aligned}
%\end{equation}
%\end{widetext}

%-----------------------bibliography------------------------------


\begin{thebibliography}{99}

\bibitem{Ficek} Z. Ficek and S. Swain, Quantum Interference and Coherence: Theory and Experiments, Springer Series in Optical Sciences Vol. 100 (Springer Science, New York, 2005).

\bibitem{Scully} M.O. Scully, M.S. Zubairy, G.S. Agarwal, and H. Walther, Science \textbf{299}, 862 (2003).
\bibitem{Quan2006} H.T. Quan, P. Zhang, and C.P. Sun, Phys. Rev. E \textbf{73}, 036122 (2006).
\bibitem{Sun} J.Q. Liao, H. Dong, and C. P. Sun, Phys. Rev. A \textbf{81}, 052121 (2010).

\bibitem{Kim} S.W. Kim and M.S. Choi, Phys. Rev. Lett. \textbf{95}, 226802 (2005).
\bibitem{Pielawa} S. Pielawa, G. Morigi, D. Vitali, and L. Davidovich, Phys. Rev. Lett. \textbf{98}, 240401 (2007).
\bibitem{Sarlette} A. Sarlette, J.M. Raimond, M. Brune, and P. Rouchon, Phys. Rev. Lett. \textbf{107}, 010402 (2011).


\bibitem{Poyatos} J.F. Poyatos, J.I. Cirac, and P. Zoller, Phys. Rev. Lett. \textbf{77}, 4728 (1996).
\bibitem{Myatt} C.J. Myatt, B.E. King, Q.A. Turchette, C.A. Sackett, D. Kielpinski, W.M. Itano, C. Monroe, and D.J. Wineland, Nature (London) \textbf{403}, 269 (2000).
\bibitem{Diehl} S. Diehl, A. Micheli, A. Kantian, B. Kraus, H. P. Buchler, and P. Zoller, Nature Phys. \textbf{4}, 878 (2008).
\bibitem{Wolf} F. Verstraete, M.M. Wolf, and J.I. Cirac, Nature Phys. \textbf{5}, 633 (2009).
\bibitem{Clerk} Y.D. Wang and A.A. Clerk, arXiv:1301.5553.

\bibitem{Engel} G.S. Engel, T.R. Calhoun, E.L. Read, T.K. Ahn, T. Man$\mathrm{\check{c}}$al, Y.C. Cheng, R.E. Blankenship, and G.R. Fleming,
Nature (London) \textbf{446}, 782 (2007).
\bibitem{Pani} G. Panitchayangkoon, D. Hayes, K.A. Fransted, J.R. Caram, E. Harel, J. Wen, R.E. Blankenship, and G.S. Engel,
Proc. Natl. Acad. Sci. U.S.A \textbf{107}, 766 (2010).
\bibitem{Huelga} S.F. Huelga and M.B. Plenio, Contemp. Phys. \textbf{54}, 181 (2013).
\bibitem{Wu1} J.L. Wu, F. Liu, J. Ma, R.J. Silbey, and J.S. Cao, J.Chem. Phys. \textbf{137}, 174111 (2012).
\bibitem{Wu2} J.L. Wu, R.J. Silbey, and J.S. Cao, Phys. Rev. Lett. \textbf{110}, 200402 (2013).
\bibitem{Kassal} I. Kassal,  J.Y. Zhou, and S.R. Keshari, J. Phys. Chem. Lett. \textbf{4}, 362 (2013).

\bibitem{Ishizaki} A. Ishizaki and  G.R. Fleming, Annu. Rev. Phys. Chem. \textbf{3}, 333 (2012).
\bibitem{Lee1} H. Lee, Y. C. Cheng, and G. R. Fleming, Science \textbf{316}, 1462 (2007).
\bibitem{Cheng1} Y. C. Cheng and R. J. Silbey, Phys. Rev. Lett. \textbf{96}, 028103 (2006).
\bibitem{Collini1} E. Collini, C. Y. Wong, K. E. Wilk, P. M. G. Curmi, P. Brumer, and G. D. Scholes,Nature \textbf{463}, 644 (2010).


\bibitem{Hormoz} S. Hormoz, Phys. Rev. E \textbf{87}, 022129 (2013).
\bibitem{Correa1}L. A. Correa, J. P. Palao, G. Adesso, and D. Alonso, Phys. Rev. E \textbf{87}, 042131 (2013).
\bibitem{Quan1} H.T. Quan, Y.X. Liu, C.P. Sun, and F. Nori, Phys. Rev. E \textbf{76}, 031105 (2007)
 \bibitem{Wang1} H. Wang, S.Q. Liu, and J.Z. He, Phys. Rev. E \textbf{79}, 041113 (2009).
\bibitem{Linden1} N. Brunner, M. Huber, N. Linden, S. Popescu, R. Silva, and P. Skrzypczyk, arXiv:1305.6009v1.
\bibitem{Grimsmo1} A.L Grimsmo, Phys. Rev. A \textbf{87}, 060302 (2013).
\bibitem{Ueda1} K. Funo, Y. Watanabe, and M. Ueda, Phys. Rev. A \textbf{88}, 052319 (2013).
\bibitem{Kim1} J.J. Park, K.H. Kim, T. Sagawa, and S.W. Kim, Phys. Rev. Lett. \textbf{111}, 230402 (2013).

\bibitem{Huang1} X. L. Huang, T. Wang, and X. X. Yi, Phys. Rev. E \textbf{86}, 051105 (2012).
\bibitem{Correa2} L. A. Correa, J. P. Palao, D. Alonso, G. Adesso, Scientific Reports. \textbf{4}, 3949 (2014).

\bibitem{Abah1} O. Abah and E. Lutz, arXiv:1303.6558v2.
\bibitem{Mehta} P. Mehta and A. Polkovnikov, Ann. Phys. \textbf{332}, 110 (2012).
\bibitem{Lutz} R. Dillenschneider and E. Lutz, Europhys. Lett. \textbf{88}, 50003 (2009).

\bibitem{Plenio} T. Baumgratz, M. Cramer, and M.B. Plenio, arXiv:1311.0275v1.

\bibitem{Scully1} M.O. Scully, Phys. Rev. Lett. \textbf{104}, 207701 (2010).
\bibitem{Scully2} M.O. Scully, K.R. Chapin, K.E. Dorfman, M.B. Kim, and A.y. Svidzinsky, Proc Natl Acad Sci USA \textbf{108}, 15097 (2011).
\bibitem{Nalbach} P. Nalbach and M. Thorwart, Proc Natl Acad Sci USA \textbf{110}, 2693 (2013).
\bibitem{Dorfman} K.E. Dorfman, D.V. Voronine, S. Mukame, and M.O. Scully, Proc Natl Acad Sci USA \textbf{110}, 2746 (2013).



\bibitem{Meystre} P. Filipowicz, J. Javanainen, and P. Meystre, Phys. Rev. A \textbf{34}, 3077 (1986).
\bibitem{Scully4} M.O. Scully and M.S. Zubairy, \emph{Quantum Optics} (Cambridge University Press, Cambridge, 1997).
\bibitem{Scully3} M.O. Scully and W.E. Lamb, Phys. Rev. \textbf{159}, 208 (1967).
\bibitem{Orszag} M. Orszag, \emph{Quantum Optics: Including Noise Reduction, Trapped Ions, Quantum Trajectories, and Decoherence} (Springer, Berlin, 2007).
\bibitem{Meschede} D. Meschede, H. Walther, and G. M\"{u}ller, Phys. Rev. Lett. \textbf{54}, 551 (1985).
\bibitem{Casagrande} F. Casagrande, M. Garavaglia, and A. Lulli, Opt. Comm. \textbf{151}, 395 (1998).
\bibitem{Cresser} J.D. Cresser, Phys. Rev. A \textbf{46}, 5913 (1992).
\bibitem{Cresser96} J.D. Cresser and S.M. Pickles, Quan. and Semiclassical Optics \textbf{8}, 73 (1996).
\bibitem{Bergou} J. Bergou, L. Davidovich, M. Orszag, C. Benkert, M. Hillery, and M.O. Scully, Phys. Rev. A \textbf{40}, 5073 (1989); J. Bergou and P. K\'{a}lm\'{a}n, Phys. Rev. A \textbf{43}, 3690 (1991).
\bibitem{Guerra} E.S. Guerra, A.Z. Khoury, L. Davidovich, and N. Zagury, Phys. Rev. A \textbf{44}, 7785 (1991).
\bibitem{Brune} M. Brune, J.M. Raimond, P. Goy, L. Davidovich, and S. Haroche, Phys. Rev. Lett. \textbf{59}, 1899 (1987).
\bibitem{Rempe} G. Rempe, F. Schmidt-Kaler, and H. Walther, Phys. Rev. Lett. \textbf{64}, 2783 (1990).

\bibitem{Wootters} W.K. Wootters, Phys. Rev. Lett. \textbf{80}, 2245 (1998).
\bibitem{Ollivier} H. Ollivier and W.H. Zurek, Phys. Rev. Lett. \textbf{88}, 017901 (2001).
\bibitem{Wang} C.Z. Wang, C.X. Li, L.Y. Nie, and J.F. Li, J. Phys. B: At. Mol. Opt. Phys. \textbf{44}, 015503 (2011).

\bibitem{Esposito} N. Golubeva, A. Imparato, and M. Esposito, Phys. Rev. E \textbf{88}, 042115 (2013).

\bibitem{Lee} W. Lee, K. Kim, W. Jeong, L.A. Zotti, F. Pauly, J.C. Cuevas,
and P. Reddy, Nature \textbf{498}, 209 (2013).


\end{thebibliography}
\end{document}